\documentclass[12pt]{article}
\usepackage{graphicx}
\usepackage{epstopdf}
\usepackage{amsmath}
\usepackage{amsfonts}
\usepackage{amssymb}
\usepackage{color}
\usepackage{mathrsfs}

\pdfoutput=1

\newcommand{\be}{\begin{equation}}
\newcommand{\ee}{\end{equation}}
\newcommand{\bea}{\begin{eqnarray}}
\newcommand{\eea}{\end{eqnarray}}
\newcommand{\barr}{\begin{array}}
\newcommand{\earr}{\end{array}}

\usepackage[colorlinks,bookmarks]{hyperref}
\definecolor{linkblue}{rgb}{0,0,0.8}
\definecolor{linkgreen}{rgb}{0,0.5,0}

\hypersetup{pdfpagemode=UseNone, pdfstartview=FitH, linkcolor=linkblue, %
            citecolor=linkgreen, urlcolor=linkblue}

\def\beq{\begin{equation}}
\def\eeq{\end{equation}}
\def\be{\begin{equation}}
\def\ee{\end{equation}}
\def\bea{\begin{eqnarray}}
\def\eea{\end{eqnarray}}
\def\d{{\partial}}

\def\nn{\nonumber}

\def\knl{{k_{\rm NL}}}
\def\km{{k_{\rm M}}}
\def\vnl{{v_{\rm NL}}}
\def\vvir{{v_{\rm vir}}}
\def\xfl{{\vec x_{\rm fl}}}
\def\epssm{{\epsilon_{s<}}}
\def\epssM{{\epsilon_{s>}}}
\def\epsdm{{\epsilon_{\delta<}}}
\def\oms{{\omega_{\rm short}}}
\def\tin{{t_{\rm in}}}

\newcommand{\ktr}{k_{\rm tr}}
 \newcommand{\tknl}{{\tilde k}_{\rm NL}}
\newcommand{\invMpc}{\,h\, {\rm Mpc}^{-1}\,}
\newcommand{\hinvMpc}{\,h\, {\rm Mpc}^{-1}\,}
\def\H{{\cal H}}

\newcommand{\vk}{\vec{k}}

\newcommand{\vq}{\vec{q}}

\begin{document}


\setcounter{page}{1} \baselineskip=15.5pt \thispagestyle{empty}

\begin{flushright}
\end{flushright}

\begin{center}

{\Large \bf Bias \\[0.5cm] in the Effective Field Theory of Large Scale Structures}
\\[0.7cm]
{\large Leonardo Senatore${}^{1,2}$ }
\\[0.7cm]
{\normalsize { \sl $^{1}$ Stanford Institute for Theoretical Physics,\\ Stanford University, Stanford, CA 94306}}\\
\vspace{.3cm}

{\normalsize { \sl $^{2}$ Kavli Institute for Particle Astrophysics and Cosmology, \\
SLAC and Stanford University, Menlo Park, CA 94025}}\\
\vspace{.3cm}

\end{center}

\vspace{.8cm}

\hrule \vspace{0.3cm}
{\small  \noindent \textbf{Abstract} \\[0.3cm]
\noindent We study how to describe collapsed objects, such as galaxies, in the context of the Effective Field Theory of  Large Scale Structures. The overdensity of galaxies at a given location and time is determined by the initial tidal tensor, velocity gradients and spatial derivatives of the regions of dark matter that, during the evolution of the universe, ended up at that given location. Similarly to what recently done for dark matter,  we show how this Lagrangian space description can be recovered by upgrading simpler Eulerian calculations. We describe the Eulerian theory. We show that it is perturbatively local in space, but non-local in time, and we explain the observational consequences of this fact.  We give an argument for why to a certain degree of accuracy the theory can be considered as quasi time-local and explain what the operator structure is in this case. We describe renormalization of the bias coefficients so that, after this and after upgrading the Eulerian calculation to a Lagrangian one, the perturbative series for galaxies correlation functions  results in a manifestly convergent expansion in powers of $k/k_{\rm NL}$ and $k/k_{\rm M}$, where $k$ is the wavenumber of interest, $k_{\rm NL}$ is the wavenumber associated to the non-linear scale, and $k_{\rm M}$ is the comoving wavenumber enclosing the mass of a galaxy.
 
 \vspace{0.3cm}
\hrule


 \vspace{0.3cm}

\section{Introduction}

The Effective Field Theory of Large Scale Structures (EFTofLSS)~\cite{Baumann:2010tm,Carrasco:2012cv,Porto:2013qua,Senatore:2014via} has been so far focussed on predicting the two-point function of dark matter fluctuations. The two-loop results~\cite{Carrasco:2013mua}, after IR-resummation~\cite{Senatore:2014via}, show a remarkable 1\% agreement with $N$-body simulations up to the very high wavenumber $k\simeq 0.6\hinvMpc$ at redshift zero. Similarly promising results have been obtained at one-loop for the momentum  power spectrum~\cite{Senatore:2014via} and the dark matter bispectrum~\cite{Angulo:2014tfa,Baldauf:2014qfa}.

It should be stressed that these results have been obtained only in the context of a very limited set of correlation function and at one redshift, even though these are still very non-trivial results with several independent checks that suggest, but do not prove, that these results are not just due to luck, but that instead dark matter physics is more understandable that previously believed. If the $k$-reach of these results were to extend unaltered to all observables in Large Scale Structure (LSS), then the picture of what we expect to be able to learn from next generation LSS surveys would change completely. Former analytic techniques stop agreeing with simulations at about  $k\simeq 0.1\hinvMpc$. Since the number of available modes grows as $k^3$, the results of the EFTofLSS tell us that there is the potential of a factor of 200 more modes available to analytical techniques than previously believed. Since error bars of optimal estimators go proportionally to the inverse the number of available modes, the consequences for what we can learn about cosmology from next generation LSS surveys can be huge.

Dark matter correlation functions are important not only because they are directly observable in lensing surveys, but also because they are the fundamental block upon which predictions for other observables, such as galaxy correlations, can be constructed. Without a correct theory of dark matter, little can be done for the rest of the observables. The purpose of this paper is to develop the formalism to construct galaxy observables from the theory of dark matter.  

Our analysis will apply to all collapsed objects such as halos and galaxies, to which we will refer interchangeably. These are thought to be biased tracers of the underlying dark matter distribution. With this we mean that if a galaxy of a certain mass will form at a given location at a given time can be traced to the distribution of dark matter at that given location. Building on important earlier works,  we will explore this relationship in great details, developing an effective parametrization of the dependence of the number density of galaxies on the underlying dark matter density. As for the theory of dark matter, we will show that the theory for collapsed objects is naturally and necessarily formulated in Lagrangian space, according to which the number density of galaxies at a given location is a function of the dark matter field evaluated at the initial location that was occupied by the dark matter fluid cell that ended up, through the evolution of the universe, at that given location. This leaves us with an undetermined function. Symmetry considerations will show that the function will depend on powers of the dark matter field, the tidal tensor, and derivatives of the velocity field, all evaluated at the initial location. After Taylor expansion, this will lead to several bias coefficients (from which the saying that galaxies are biased tracers of the underlying dark matter distribution).  Since the formation of galaxies at a given point does not depend on the dark matter only exactly at the same location, but also at neighboring locations, we will allow also for dependence on spatial derivatives of these fields. The scale suppressing these derivatives is the comoving scale associated to the mass of the given galaxy. We call this $\km$. This scale is different in general from the so-called non-linear scale $\knl$ that suppressed the higher derivative terms as well as the loop corrections in the dark matter correlation functions.
 
As described in the case of dark matter~\cite{Senatore:2014via}, calculations in Lagrangian space create some practical complications with  renormalization. In practical terms, it is much easier to extend Eulerian calculations to Lagrangian calculations by resumming the large scale IR displacement fields. This was recently done in~\cite{Senatore:2014via} for dark matter, and we generalize it here for galaxies.  This leads us to develop the Eulerian theory for collapsed objects. Here the galaxy density at a given location is a function of the dark matter, tidal tensor, and velocity fields, and their spatial derivatives, all evaluated on the past trajectory of the fluid element that ended up at a given location. This formulation is sensitive to IR contributions that are not under control, which make it necessary to upgrade the calculation to a Lagrangian one, as we just described. This is also a peculiar theory, as it is non-local in time. We show how precisely to deal with this peculiar feature and work out some of the observational consequences. We also give an argument for why, to some degree of approximation, the Eulerian EFTofLSS can be considered as quasi local in time. In this case, the field of collapsed objects must depend also on the time derivative along the flow of the tidal tensor and velocity gradients.

We give very explicit formulas for the dark matter galaxies cross correlation functions at one loop, and almost as explicit for the galaxies galaxies power spectrum. Similar formulas can be easily derived for higher legs and higher order correlation functions. The example of the dark matter galaxies cross correlation allows us to illustrate how, following in part~\cite{McDonald:2009dh}, under a perturbative calculations, the bias coefficients need to be renormalized order by order in perturbation theory, so that the final result is independent of the short distance physics that naively contributes inside the loop integrals. 

After the renormalization is performed, the loop expansion, completed by the insertion of the relevant higher order bias coefficients, amounts to an expansion in the parameters that control the dark matter expansion: $\epsilon_{\delta<}$ and $\epsilon_{s>}$~\cite{Porto:2013qua,Senatore:2014via}. These are defined as 
\bea
&&\epsilon_{s >} =k^2  \int_k^\infty {d^3k' \over (2 \pi)^3}  {P_{11}(k') \over k'^2}\ , \\ \nonumber
&&\epsilon_{\delta <} = \int^k_0 {d^3k' \over (2 \pi)^3} P_{11}(k')\ .
\eea
where $P_{11}(k)$ is the dark matter power spectrum. $\epsilon_{s >} $ represents the displacement due to short wavelength modes, while $\epsilon_{\delta <}$ represents the tidal force due to long wavelength modes.  Both of these scale  proportionally to $k/\knl$. The bias derivative expansion corresponds to an expansion in powers of $(k/\km)^2$. Finally, in the Eulerian treatment we expand also in displacement due to long wavelength modes $\epsilon_{s<}= (k\, \delta s_<)^2$, where
\bea
\epsilon_{s_<} &=&k^2  \int_0^k {d^3k' \over (2 \pi)^3}  {P_{11}(k') \over k'^2}\ .
\eea
As described in~\cite{Senatore:2014via}, $\epsilon_{s<}$ is of order one for the $k$'s of interest, and therefore one cannot Taylor expand in this parameter. This forces us to pass to the Lagrangian description, which does not expand in $\epsilon_{s<}$ but only in $\epsilon_{s>}$ and $\epsilon_{\delta<}$. This means that at this point we establish a perturbative expansion in which each successive perturbative order scales as an higher powers of $k/\knl\ll1$ and $k/\km\ll1$. This is a manifestly convergent expansion, that will converge to the true answer until non-perturbative effects become important at a wavenumber near $\knl$.

This paper develops the relevant equations that are necessary to make predictions for correlation functions involving galaxies and collapsed objects. We leave to future work to compare the resulting predictions with the simulated data.

\section{Eulerian Description and its incompleteness}

We wish to write how the distribution of galaxies depends on the distribution of the dark matter.  Galaxies form because of gravitational collapse, therefore they will depend on the underlying values of the gravitational field and dark matter field.  Since the overdensities of galaxies is a scalar quantity, it can only depend on similarly scalar quantities built out of these fields.
 Let us consider each of these terms one at a time. 

Concerning the gravitational field, because of the equivalence principle, the number of galaxies at a given location can only depend on the gravitational potential $\phi$ with at least two derivatives acting on it, as it is for the curvature.  $\phi$ without derivatives does  appear in curvature terms only at non-linear level in terms such as $\phi \d^2\phi$ or $(\d\phi)^2$.  These are general relativistic corrections, which are important only at long distances of  order Hubble, where perturbations can be treated as linear to a very good approximation. We will therefore neglect these terms.

In the Eulerian EFT, the dark matter field is identified by the density field $\delta$ and the momentum field $\pi^i$~\cite{Carrasco:2013mua}. This is a useful quantity because its divergence is related to the time derivative of the matter overdensity by the continuity equation.  Due to Newton's equation, the density field is constrained to be proportional to $\d^2\phi$, so it can be discarded as an independent field. Concerning the momentum field, clearly a spatially constant momentum field cannot affect the formation of galaxies. Indeed, the momentum is not a scalar quantity. Under a spatial diffeomorphism
\be
x^i\quad\to\quad x^i+\int^\tau d\tau'\; V^i
\ee 
the momentum shifts as
\be
\pi^i\quad\to\quad \pi^i + V^i \rho\ .
\ee 
where $\rho$ is the dark matter density $\rho=\rho_b (1+\delta)$, with $\rho_b$ being the background density.
Similarly, the gradient of the momentum shifts as
\be
\d_j\pi^i\quad\to\quad \d_i\pi^i + V^i \rho_b \d_j\delta
\ee 
We can form a scalar quantity by combining $\d_j\pi^i$ with $v^i \d_j\delta$:
\be
[\bar \d_j\pi^i]\equiv \d_i\pi^j- v^j \rho_b \d_i\delta\ ,
\ee
which trivially transforms as a scalar.

Working with the field $\pi^i$ has the advantage, as discussed in~\cite{Carrasco:2013mua}, that no new counterterm is needed to define correlation functions of $\d_i\pi^i$ once the correlation functions of $\delta$ have been renomalized. Alternatively, one can work with the velocity field $v^i$, defined as
\be
v(\vec x,t)^i=\frac{\pi(\vec x,t)^i}{\rho(\vec x,t)} \ . 
\ee
The velocity field has the advantage that $\d_iv^j$ is a scalar quantity. However, $v^i$ is defined as the ratio of two operators at the same location. It is therefore a composite operator that requires its own counterterm and a new renormalization even after the matter correlation functions have been renormalized~\cite{Carrasco:2013mua}. As we will see, when dealing with biased tracer, one has to define contact  operators in any event, and $v^i$ has simpler transformation properties than $\pi^i$. Therefore, instead of working with $\pi^i$, we work with $v^i$. In analogy to what we have just discussed, the galaxy field can depend on $v^i$ only through $\d_j v^i$ and its derivatives. 

The field of collapsed objects at a given location will not depend just on the gravitational field or the derivatives of the velocity field at the same location. There will be a length scale enclosing the points of influence. This length scale will be of order the spatial range covered by the matter that ended up collapsing in a given collapsed object. We call the wavenumber associated to this scale $k_{M}$, as it depends on the nature of the object, most probably prominently through its mass. We expect $\km\sim 2\pi (\frac{4\pi}{3}\frac{\rho_{b,0}}{M})^{1/3}$, where $M$ is the mass of the object and $\rho_{b,0}$ is the present day matter density. In particular, $\km$ can be different from $\knl$, the scale as which the dark matter field becomes non-linear~\footnote{$\knl$ can be unambiguously defined as the scale at which dark matter correlation functions  computed with the EFT stop converging.}. If we are interested on correlations on collapsed objects of wavenumbers $k\ll \km$, we can clearly Taylor expand this spatially non-local dependence in spatial derivatives. 
 
 In addition, in general there is a difference between the average dependence of the galactic field on a given realization of the long wavelength dark matter fields, and its actual response in a specific realization. To account for this, we add a stochastic term $\epsilon$ to the general dependence of the galaxy field.   $\epsilon$ is a stochastic variable with zero mean but with other non-trivial correlation functions.
    
In summary, we are lead to an expression for the dependence of the number density of galaxies of kind $M$ at position $\vec x$ and time $t$ of the following form:
\be\label{eq:euler_bias_0}
\delta_M(\vec x,t)= f\left(\d_i\d_j \phi(\vec x,t),\d_j v^i(\vec x,t), \frac{\d^i}{\km},\epsilon(\vec x,t)\right)\ ,
\ee
where $f$ is a scalar function built with its arguments in such a way that if all the arguments vanish, then $f=0$.
Since we are interested only in long wavelength perturbations, for which all the fluctuations are smaller than one, we can Taylor expand, and define bias coefficients as the coefficients $c_i$ of this Taylor expansion
\bea\label{eq:euler_bias_1}
&&\delta_M(\vec x,t)\simeq  \\  \nonumber
&&\qquad \simeq c_{\d^2\phi}(t)\; \frac{\d^2\phi(\vec x,t)}{H^2}+ c_{\d_i v^i}(t) \;  \frac{\d_i v^i(\vec x,t)}{H}+ c_{\d_i \d_j \phi \d^i \d^j \phi}(t) \;\frac{\d_i\d_j \phi(\vec x,t)}{H^2}\frac{ \d^i \d^j \phi(\vec x,t)}{H^2} + \ldots\\\nonumber
&&\qquad+ c_{\epsilon}(t)\;\epsilon(\vec x,t)+ c_{\epsilon\d^2\phi}(t) \;\epsilon(\vec x,t)\frac{\d^2\phi(\vec x,t)}{H^2}+ \ldots \\ \nonumber
&&\qquad+  c_{\d^4\phi}(t)   \;\frac{\d^2}{\km^2}\frac{\d^2\phi(\vec x, t)}{H^2}+\dots\ ,
\eea
where in the first line we have expanded in powers of the long wavelength dark matter fluctuations, in the second we have included powers of the stochastic fluctuations, and in the third we have started including higher derivative terms. The factors of $H$ suppressing $\d^2\phi$ and $\d_i v^i$  can be inserted by rescaling the bias coefficient. The choice we made has the following useful property. We expect that if the long mode is close to the non-linear scale, than the Taylor expansion should simply not-converge. Because of this, if we take the bias coefficients of order one, then it better be that the fluctuation fields are of order one at the non linear scale. The factors of $H$ that we inserted realize this~\footnote{Apart for the distinction between the momentum field and the velocity field, all of the above results, and in particular~(\ref{eq:euler_bias_1}), were obtained in~\cite{McDonald:2009dh}. By this we mean the identification of the fields that appear in the bias expansion, of the presence of a stochastic term, and of the fact that the higher derivative terms are suppressed by a scale $\km$, which is in general different than $\knl$ (see also~\cite{Smith:2006ne,Matsubara:1999qq,Matsubara:2011ck}, and~\cite{Kehagias:2013rpa} for a discussion of the time-dependence of the bias coefficients).

There are also other important results already obtained in~\cite{McDonald:2009dh}, but that will appear later in the discussion. They are the fact that one can use the dark matter equations of motion to reduce the number of coefficients in~(\ref{eq:euler_bias_1}) and the fact that the bias coefficients $c_i$ represent the {\it bare} bias coefficients, that need to be renormalized. In~\cite{McDonald:2009dh}, the authors indeed carefully explained how to perform the renormalization of the biases. The only aspect in which so far our treatment is different from the one in~\cite{McDonald:2009dh} is that we treat the stochastic noise $\epsilon$ has a non-Gaussian variable, while it was considered Gaussian in~\cite{McDonald:2009dh}. Other important differences will soon emerge.}.

However, we are going to argue that in our opinion the former equations are incomplete.

Indeed, in the former equation~(\ref{eq:euler_bias_1}), the presence of time-derivatives is missing: only spatial derivatives appear. This is inconsistent to us. The field of collapsed objects at a given time will be sensitive not just on the gravitational field and velocity field at the same time, but also at earlier times. If the time scale $1/\oms$ associated to the short modes is much shorter than the one associated to the long ones, $1/\omega_{\rm long}$, the short modes are affected only by the long modes around a given time. Therefore, the way the long modes at earlier times affect the short modes can be efficiently parametrized by an expansion in time derivatives, such as for example $\frac{1}{\oms}\frac{\d \delta}{\d t}$, whose relative contribution scale as $\frac{\omega_{\rm long}}{\oms}$. In the perturbative treatment, one could naively imagine that these terms are degenerate with the former ones, because the time dependence of the modes in perturbation theory is just given by the growth factor and therefore is $k$-independent, where $k$ is the wavenumber of a mode.  This is  however misleading for two reasons. First, these terms are suppressed by powers of $\oms$, which is in general different than $1/\knl$, and so they appear specifically in the power counting. Second, the degeneracy between $H \delta$ and $\dot\delta$ is only true at linear level, but fails at non-linear level. To realize  this, just notice that $\langle\delta(\vec k,t)\delta(\vec k,t)\rangle$ is IR-safe, while  $\langle\dot\delta(\vec k,t)\dot\delta(\vec k,t)\rangle$ is not IR-safe. 

This suggest that we should add in the bias terms  that go as $\frac{1}{\oms}\frac{\d }{\d t}$, such as $\frac{1}{\oms}\frac{\d\; \d^2\phi }{\d t}$.  It is pretty clear that these term are not diff. invariant.  Under a time-dependent spatial diff., $\d/\d t$ shifts as~\footnote{People familiar with the Effective Field Theory of Inflation~\cite{Cheung:2007st,Senatore:2010wk} might remember that $g^{0\mu}\d_\mu$ is invariant, not $\d/\d t$.}
\be
\frac{\d}{\d t}\quad\to\quad \frac{\d}{\d t}- V^i \frac{\d}{\d x^i}\ .
\ee
A diff. invariant combination can be formed by allowing the presence of the dark matter velocity field $v^i$ without derivatives acting on it, and defining a {\it flow time-derivative}, familiar from fluid dynamics, as
\be
\frac{D}{D t}=\frac{\d}{\d t}+ v^i \frac{d}{d x^i}\ .
\ee
We are therefore led to naively lead to include terms of the form
\be
\delta_M(\vec x, t)\supset c_{D_t \d^2\phi}(t)\;\frac{1}{H^2}\frac{1}{\oms}\frac{D\,\d^2 \phi }{D t} + \ldots\  .
\ee

In reality, the situation is even more peculiar, at least at first. In fact, let us ask ourselves what is the scale $\oms$ that suppresses the higher derivative operators. Naively, $\oms$ is of order $H$, as this is the timescale of the short modes collapsing into halos. This is the same time-scale as the long modes we are keeping in in our effective theory! This means that the parameters controlling the Taylor expansion in $\frac{1}{\oms}\frac{D}{D t}\sim \frac{H}{\oms}$ is actually of order one. Therefore, what we have to do is to generalize these formulas: since the formation time of a collapsed object is of order Hubble, we have to allow for the density of the collapsed objects to depend on the underlying long-wavelength fields evaluated at all times up to an order one Hubble time earlier. This means that the formula relating compact objects and long-wavelength fields will actually be non-local in  time. Therefore we have
\bea\label{eq:euler_bias_2}
&&\delta_M(\vec x,t)\simeq \int^t dt'\; H(t')\; \left[   \bar c_{\d^2\phi}(t,t')\; \frac{\d^2\phi(\xfl,t')}{H(t')^2} \right.\\  \nonumber
&&\quad+ \bar c_{\d_i v^i}(t,t') \;  \frac{\d_i v^i(\xfl,t')}{H(t')}+\bar c_{\d_i \d_j \phi \d^i \d^j \phi}(t,t') \;\frac{\d_i\d_j \phi(\xfl, t')}{H(t')^2}\frac{ \d^i \d^j \phi(\xfl,t')}{H(t')^2} + \ldots\\\nonumber
&&\quad+ \bar c_{\epsilon}(t,t')\;\epsilon(\xfl,t')+\bar c_{\epsilon\d^2\phi}(t,t') \;\epsilon(\xfl,t')\frac{\d^2\phi(\xfl,t')}{H(t')^2}+ \ldots \\ \nonumber
&&\left.\quad+  \bar c_{\d^4\phi}(t,t')   \;\frac{\d^2_{x_{\rm fl}}}{\km^2}\frac{\d^2\phi(\xfl,t')}{H(t')^2}+\dots\ \right] .
\eea
Here $\bar c_{\ldots}(t,t')$ are dimensionless kernels with support of order one Hubble time and with size of order one, and $\xfl$ is defined iteratively as
\beq
\xfl(\vec x, \tau,\tau') = \vec x - \int_{\tau'}^{\tau} d\tau''\; \vec{v}(\tau'',\vec x_{\rm fl}(\vec x, \tau,\tau''))\ .
\eeq
where $\tau$ is conformal time.
Notice that the higher spatial derivative is with respect to $x_{\rm fl}$, not to $x$. This makes a lot of sense, at least to us. The way a collapsed objects forms is by being affected by the distribution of matter in a region of the size of the matter that will indeed eventually collapse in the object. This length scale is $1/\km$. Therefore, the derivative expansion that perturbatively reconstructs  the distribution of matter in a region of order $1/\km$ starting from the origin should be controlled by the parameter $\km$.  However, in the long formation time of order $H^{-1}$ that it takes for an object to form, the object will have moved by a distance of order $\sqrt{\epsilon_{s}/k^2}$. Therefore, it will be affected by matter fields that in fixed  comoving-FRW coordinates will span a region of order $\sqrt{\epsilon_{s}/k^2}$. However, once going to comoving coordinates that move with a region, it is only points a distance of order $1/\km$ away that affect the collapsing of the object. This is why, if we make the spatial derivatives act along $\xfl$, these are suppressed by powers of $\km$.  In formulas, by applying the chain rule, we have the following expression
\be
\frac{\d}{\d x_{\rm fl}^i} \phi(\xfl)=\left.\frac{\d}{\d x^j} \phi(\vec x)\right|_{\vec x(\xfl))}\cdot \left.\frac{\d x^j}{\d x_{\rm fl}^i}\right|_{\vec x(\xfl))}\simeq \left.\frac{\d}{\d x^j} \phi(\vec x)\right|_{\vec x(\xfl)}\cdot \left(\delta^{j}{}_i+  \int_{\tau'}^{\tau} d\tau''\; \left.\frac{\d}{\d x^i} v^j(\tau'',\vec x)\right|_{\vec x(\xfl)}\ \right)\ .
\ee

The fact the theory for biased objects is non-local in time derives from the same logic that made us conclude that the theory is non-local in time also when describing dark matter~\cite{Carrasco:2013mua,Carroll:2013oxa}. In that case, as in this case, there is no hierarchy of time scales between the motions of dark matter particles around the non-linear scale, and the motion at much larger distances. As in the case for dark matter, therefore, the treatment will be extremely similar.  As we will explain next, both in the case of dark matter and for the collapsed objects, the solution organizes itself in a perturbative expansion where each term contributes as higher powers of the small parameters $k/\knl\ll 1$ or $k/\km\ll 1$.  Such a small parameter does not exist for the time-derivative terms, and this is why the theory is non-local in time.

There exist a simplification that can be done at an approximate level. We are now going to argue that there is an heuristic argument that allow us to infer that the theory both for dark matter and for the collapsed objects can be treated as quasi local in time, to some degree of approximation. By this we mean that the hierarchy in time derivatives can be thought of as approximately controlled by a small parameter,  $\frac{H}{\oms}$, which is actually smaller than one. This parameter is a {\it quantitatively different} parameter than $k/\knl\ll 1$ and $k/\km\ll 1$ that control the expansion in spatial derivatives. More importantly, it is {\it qualitatively different}: contrary to the expansion in $k/\knl\ll 1$ and $k/\km\ll 1$, which is very rigorous, the argument for the expansion in $H/\oms$ is approximate. Therefore, while the expansion is $k/\knl\ll 1$ and $k/\km\ll 1$ converges to the true answer~\footnote{In the sense of asymptotic series.}, the one in $H/\oms$ is different: it is only an approximation which will get close to the true answer, but not arbitrarily. As we will see shortly, the argument that leads to such an approximation is very heuristic: how good this approximation is should be carefully verified in calculations and simulations, but it is potentially very interesting. We leave this to future work.

Before proceeding to give an argument for the quasi locality in time, let us give the formulas for the velocity of the collapsed objects. The same logic that led us to express the overdensity of compact object in terms of the long wavelength dark matter and gravitational fields allows us to derive similar formulas for the momentum and the velocity of the collapsed objects. 

Under a Lorentz boost, $x^i\to x^i +V^i t$, the momentum of some objects transforms proportionally to $V^i$ and the density of the objects themselves $\rho$. On the other hand, the velocity of any field shifts simply proportionally to $V^i$. It is therefore simpler to work directly with the velocity fields. We have:
 \bea\label{eq:euler_bias_2_velocity} \nonumber
&&v^i_M(\vec x,t)\simeq v^i(\vec x,t) + \int^t dt'\;H(t')\; \left[ \bar d_{\d \d^2\phi,1}(t,t') \;  \frac{\d_{x_{\rm fl}^i}}{\km^2}\frac{\d^2\phi(\xfl,t')}{H(t')}+  \bar d_{\d^2v,1}(t,t') \;  \frac{\d_{x_{\rm fl}^i}}{\km^2}\d_jv^j(\xfl,t')\right.\\  \nonumber
&&\quad\quad\quad\quad\quad\quad\quad\quad\quad\quad+ \bar d_{\d(\d^2\phi)^2,1}(t,t') \; \frac{\d_{x_{\rm fl}^i}}{\km^2} \frac{\d^j\d_m\phi(\xfl,t')}{H(t')}\frac{\d_j\d^m\phi(\xfl,t')}{H(t')^2}+ \ldots\\\nonumber
&&\quad\quad\quad\quad\quad\quad\quad\quad\quad\quad+ \bar d_{\epsilon}(t,t')\;\epsilon^i(\xfl,t')+ d_{\epsilon\d^2\phi}(t,t') \;\epsilon^i(\xfl,t')\frac{\d^2\phi(\xfl,t')}{H(t')^2}+ \ldots \\
&&\left.\quad\quad\quad\quad\quad\quad\quad\quad\quad\quad+  \bar d_{\d^2 v\d^2\phi}(t,t')   \; \frac{\d^2\phi(\xfl,t')}{H(t')} \;\frac{\d^2_{x_{\rm fl}}}{\km^2} v^i(\xfl,t')+\dots\ \right] .
\eea
Notice that the transformation  under boosts forces the absence of a relative coefficient between  $v^i_M$ and $ v^i$. This can also be realized by imagining to go to the inertial frame comoving with the dark matter particles, write down the halo velocity in that frame, and come back to the original frame.
The momentum field can be constructed from the velocity field as
\be
\pi^i_M(\vec x,t)=\rho_M(\vec x,t) v^i_M(\vec x,t)+ \int^t dt'\; \bar e_v(t,t')\;  v^i_M(\xfl,t')+\ldots
\ee
where the terms in $\bar e_{\dots}$ represent the counterterms necessary to pass from the velocity field to the momentum field. The momentum field, being here defined as a local product of two longwavelenth fields, needs to be renormalized independently. See~\cite{Carrasco:2013mua} for a discussion about contact operators and their renormalization in the context of the EFTofLSS.

\subsection{An argument for approximate time locality\label{sec:quasi-local}}

Let us start with the description of dark matter clustering, which is also non-local in time~\cite{Carrasco:2013mua,Carroll:2013oxa}. We are going to argue that an approximate time-locality exist also in this case. As we will explain more in detail later, here the difference between locality and non-locality appears only starting at two loops, because at linear level in the counterterms the non-locality can be reabsorbed into a redefinition of the local in time counterterms~\cite{Carrasco:2013mua}. In~\cite{Carrasco:2013mua} it was also noticed that the structure of the time-dependence in perturbation theory implies that at each order in perturbation theory the non-locality in time can be accounted for by allowing for a different value of a given counterterm. It has been experimentally verified by a direct calculation that the approximation in which we make the counterterms local in time agrees more with the data than the non-local one~\cite{Carrasco:2013mua}. We are now going to give an heuristic argument for why this is the case. The reason why the EFT is non-local in time is because the modes at around the non-linear scale evolve on a time scale of order $H$, which is the same as the one of the long wavelength modes. Let us give a closer look at this statement. It is certainly true that modes for which $\delta\rho/\rho$ is of order one, their time scale is of order $H$.  This is true because, at the non linear scale where $\delta\sim 1$, we have, by the continuity equation 
\be
\d_i v^i\sim \dot\delta\ , \quad\Rightarrow\quad \vnl^i\sim \frac{H}{\knl}\ .
\ee
Therefore, to move a distance of order of the non-linear scale, it takes a time of order
\be
t_{\rm NL}\sim \frac{1}{\knl \vnl}\sim \frac{1}{H}\ ,
\ee
as we wished to verify.

However, as the modes keep collapsing and become more non-linear, $\delta\rho/\rho$ grows, and the modes become faster. We cannot use anymore the linear approximation, but we can use a Newtonian counting. When objects virtualize, $\delta_{\rm vir}\sim 200$, and, by mass conservation, $k_{\rm vir}\sim 200^{1/3} \knl$ . Therefore, from the Poisson equation, we have
\be
\nabla^2\Phi=H^2\delta\ ,\quad\Rightarrow\quad \Phi_{\rm vir}\sim 200^{1/3} \frac{H^2}{\knl^2} \ .
\ee
For virialized structures we have $\vvir^2\sim \Phi_{\rm vir}$, and therefore the time scale associated to the virialization scale is
\be
t_{\rm vir}\sim \frac{1}{k_{\rm vir} \vvir}\sim \frac{1}{H}\frac{1}{200^{1/2}}\ll \frac{1}{H}\ .
\ee
The time scale for virialized objects is about 14 times faster than Hubble. Supposedly, modes that have just become non-linear have a time-scale of order $H$, and as modes become shorter and shorter, the associated time scale becomes faster and faster. Therefore, it is only those modes in Fourier space that are around the non-linear scale that are the ones for which the derivative expansion in time-derivatives is not applicable. One might wonder how much that shell in phase space contributes. In the limit in which it does not contribute very much, than the time-locality, in the sense of a derivative expansion, can be a good approximation. 

Let us elaborate on this. In the EFTofLSS there is a powerful, and rather intuitive, theorem that tells us that virialized structures do not contribute to the renormalization of the parameters of the EFT~\cite{Baumann:2010tm}. This means that the short modes that contribute at long distances are only the ones from the non-linear scale up to the virialization scale. If we imagine that all modes from the non-linear scale to the virialization scale contribute equally, and we estimate the time-scale associated with each mode with the relationship $v^2\sim\Phi$, then we find that the modes around the virialization scale contribute the most. This suggest that an expansion if $H/\oms$, where $\oms\sim  H/14$, is a good approximation. 

It should be stressed a very important remark. Contrary to the expansion in $k/\knl$ or $k/\km$, that can reach arbitrary precision until when non-perturbative effects become relevant, the expansion in $H/\oms\sim 1/14$ is wrong: there is an irreducible mistake that cannot be recovered by going to higher orders in $\tfrac{D_t}{\oms}$. This is the contribution from the modes around the non-linear scale.   That contribution, no matter how small, cannot be recovered with the time-derivative expansion. In summary, the expansion in $\tfrac{D_t}{\oms}$ is an approximate expansion that allow us to approximate the theory as local in time for the first orders in perturbation theory.  Naive phase space arguments suggest that the contribution of the non-linear modes is about $1/200$ of the ones at the virialization scale, but this is clearly an underestimate: first the modes at the virialization scale stop contributing by the non-renormalization theorem, so it is actually modes a bit more slower that contribute the most; second, modes in between the virialization scale and the non-linear scale contribute more in phase space than $1/200$, and they move on a  time scale in between the one of the virialized modes and the one of the ones at the non-linear scale. This suggest that the size of the irreducible mistake done when expanding in $\tfrac{D_t}{\oms}$ is larger than 1/200, but probably less than one. It would be interesting to explore the size of this irreducible, systematic, mistake, and we leave this to future work. For this paper, when dealing with the approximate time-local treatment, we take $H/\oms\sim 1/10$, and we assume the size of the systematic mistake to be smaller than this.

In~\cite{Carrasco:2013mua}, for the dark matter power spectrum evaluated at two loops,  it was found that the local approximation seems to be a very good fit to the data. The reason is now clear. In this quantity, the difference between local in time and non-local in time appears first in the term where the $c_s$ counterterm is evaluated at one loop. We explain this in detail in~App.~\ref{app:fluid_bias}, subsection~\ref{app:approx_time}. This term is of the order, and actually a bit smaller, than a two loop term, and therefore it is the highest order term included in the calculation. It therefore makes sense to evaluate it in an approximate way, and our argument suggest that the error in performing a local approximation in the full evaluation of the two-loop power spectrum is about $10\%$ times a two loop term, which is about the size of a three loop term, and therefore negligible.   

The same discussion trivially extends to the collapsed object. Indeed, it is precisely the dark matter virialized structures that we tend to identify as dark matter halos. It is therefore clear that for all collapsed object such an approximate quasi local-in-time treatment is doable, with a similar irreducible mistake.

In this case, the {\it approximate} equation for the collapsed objects, where approximate is to distinguish it from (\ref{eq:euler_bias_2}), which is perturbatively exact,  becomes
\bea\label{eq:euler_bias_3}
&&\delta_M(\vec x,t)\simeq  \\  \nonumber
&&\qquad\simeq c_{\d^2\phi}(t)\; \frac{\d^2\phi(\vec x,t)}{H^2}+ c_{\d_i v^i}(t) \;  \frac{\d_i v^i(\vec x,t)}{\oms}+ c_{\d_i \d_j \phi \d^i \d^j \phi}(t)\;\frac{\d_i\d_j \phi(\vec x,t)}{H^2}\frac{ \d^i \d^j \phi(\vec x,t)}{H^2} + \ldots\\\nonumber
&&\qquad+ c_{\epsilon}(t)\;\epsilon(\vec x,t)+ c_{\epsilon\d^2\phi}(t) \;\epsilon(\vec x,t)\frac{\d^2\phi(\vec x,t)}{H^2}+ \ldots \\ \nonumber 
&&\qquad+  c_{\d^4\phi}(t)   \;\frac{\d^2}{\km^2}\frac{\d^2\phi(\vec x, t)}{H^2}
\eea
\bea \nonumber
&&\quad\quad\ + \frac{1}{H^2\,\km^2\,\oms}c_{D_t\d^4\phi}(t)  \left[-\d^2 v^i(\vec x,t) \d_i\d^2\phi(\vec x,t)+2 \d_i v^j(\vec x,t) \d_j\d_i\d^2\phi(\vec x,t)+\frac{D}{Dt}\d^2\d^2\phi(\vec x,t) \right]\\ \nonumber 
&&\qquad  + c_{D_t \d^2\phi}(t) \; \frac{1}{H^2 \oms }\frac{D}{Dt} \d^2\phi(\vec x,t)+ c_{D_t\d_i v^i}(t) \;  \frac{1}{\oms^2}\frac{D}{Dt }\d_i v^i(\vec x,t)+\ldots\ .
\eea

The terms in the third lines represent the quasi-local description of the spatial derivatives with respect to $\xfl$. Notice that, by units, the velocity terms are suppressed by the short time scale $\oms$~\footnote{This consideration does not apply to the terms in $\d^2\phi/H^2$, as they are related to matter by a constraint equation that involves $H$.}. If this scale becomes of order $H$, then the theory is effectively non-local in time, as indeed in the former section. Exactly the same consideration applies to the scale suppressing the along-the-flow time-derivative~$D_t$. It should be stressed that the argument we provided for approximate time-locality is quite heuristic and preliminary, better studies and verifications should be performed. Therefore, an equivalent way to summarize this discussion is to say that there is the opportunity for the time derivatives and the velocity gradients to be suppressed by a scale faster than Hubble. If this will turn out to be the case, than an approximate time-local treatment will be possible.   \\

The result derived so far, both in the local in time approximation, as well as in the more correct non-local treatment, present a problem. In the Eulerian treatment, the expansion parameter of perturbation theory is not just $\epsdm$, $\epssM$ or $k/\km$, as we would hope, but there is another expansion parameter, $\epssm$. For IR-safe quantities, expanding in $\epssm$ implies that the Baryon Acoustic Oscillations~(BAO) cannot be reconstructed correctly~\cite{Senatore:2014via}. For non IR-safe quantities, such as the non-equal-time matter power spectrum, the expansion in  $\epssm$ implies that perturbation theory breaks at very low $k$'s. These issues have been recently physically explained and addressed first in~\cite{Porto:2013qua}, where a Lagrangian formulation of the EFTofLSS, that is a formulation where one is not expanding in $\epssm$, has been developed, and then in~\cite{Senatore:2014via}, where the issue of the BAO oscillations and an actual very simple implementation of the Lagrangian-space calculations from the Eulerian calculations has been developed and presented for the dark matter clustering.

Exactly the same issues are present when dealing with collapsed objects. This is a serious problem and incompleteness for the Eulerian treatment, which makes it ultimately non-convergent to the true answer. For this reason, we move to a Lagrangian formulation of the bias. 

\section{Lagrangian Description\label{sec:lagrangian}}

In the Lagrangian description of the EFTofLSS when applied to dark matter, non-linear regions are thought of as extended objects endowed with mass, quadrupole, etc., moving under gravity and sourcing gravity through their overall energy~\cite{Porto:2013qua}. Let us now pass to the collapsed objects. Given a certain realization of the universe, the evolution of the density fluctuations and of the collapsed objects is deterministic. Therefore, given the initial value of the density, velocity and gravitational fields, one can in principle determine where, when, and in which size a certain collapsed object will form. When and in which size it will form will depend on the same variables that entered in the Eulerian description, such as $\d^2\phi$ and $\d_i v^j$. Where it will form will be simply given by the following: given an initial location of an extended object labelled by $\vec q$, where there is a certain value of $\d^2\phi, \d_i v^j,\ldots$, that will lead to a certain amount of collapsed objects, these objects will find themselves at the final location $\vec z(\vec q,t)$ where the initial extended region has ended up. This leads to the following formula for the density $\rho_M$ and overdensity $\delta_M$ of collapsed objects of kind $M$, where $M$ stays for the characteristics that identify the collapsed objects of interest, such as mass, color, shape, etc.~\footnote{The Lagrangian approach to biased objects has been developed for quite some time in, for example,~\cite{Matsubara:2008wx,Matsubara:2011ck,Carlson:2012bu,Biagetti:2014pha}. However, to our knowledge, the complete list of relevant operators has never been introduced.}:
\bea\label{eq:bias_lagrangian_1}
&&\rho_M(\vec x,t)=\int d^3 q\; \bar\rho_M(t)\; \delta^{(3)}(\vec x-\vec z(\vec q,t))\;\\ \nonumber
&&\qquad\qquad  \qquad \times \exp\left[f_M(\d_i\d_j\phi(\vec z(\vec q,\tin),\tin),\d_j v^i(\vec z(\vec q,\tin),\tin),\frac{\d}{\km},\epsilon(\vec z(\vec q,\tin),\tin),t)\right]\ .
\eea
Here  $\bar\rho_{M}(t)$ is the unperturbed density of collapsed objects at time $t$, while $\bar\rho_M(t) \exp[f_M(\vec q,\tin,t)]\equiv\rho_L(\vec q,\tin, t)$ is the number  of locations per unit cell $d^3q$  at time $\tin$ where collapsed objects of that specific kind will form at time $t$. We now write $\rho_M(\vec x,t)=\bar\rho_{M}(t)(1+\delta_M(\vec x,t))$, so that we are led to
\bea\label{eq:delta_galaxy_1b}
&&1+\delta_M(\vec x,t)=\\ \nonumber
&&\quad \int d^3 q\; \delta^{(3)}(\vec x-\vec z(\vec q,t))\; \exp\left[f_M(\d_i\d_j\phi(\vec z(\vec q,\tin),\tin),\d_j v^i(\vec z(\vec q,\tin),\tin),\frac{\d}{\km},\epsilon(\vec z(\vec q,\tin),\tin),t)\right]\ .
\eea
Let us explain this formula in some detail, as it is crucial for our approach. $f_M$ is a scalar function of its arguments that vanishes if the arguments vanish. The exponentiation is introduced for notational convenience, it is actually irrelevant for the actual computations, as we will see. The overdensity of collapsed objects at a location $\vec x$ and at time $t$ is a function of the overdensity, gravitational fields, etc., evaluated at some initial time and a location given by the initial location of the extended objects that end up at $\vec x$ at time $t$.

One might wonder why the function $f_M$ depends only of its arguments evaluated at the same initial time; that is, why it does not depend on its variables in a non-local in time way. In principle, one can take the initial time early enough so that all modes of interest are outside of the horizon and are constant in time. In this case the function $f_M$ would be written just in terms of the usual variable $\zeta$ that is constant at all orders in perturbation theory when outside of the horizon~\cite{Senatore:2012ya}. Much more simply, one can take the initial time to be early enough so that the non-linear corrections are completely negligible and no collapsed objects have formed. In this case, the status of the universe is completely described by the long wavelength fields at that time, without need to talk about earlier times. This implicitly defines the function $f_M$~\footnote{In fact, using the notation of the Lagrangian-space EFT that we introduce later in this section, eq.~(\ref{eq:delta_galaxy_1b}) can be rewritten as
\bea\label{eq:delta_galaxy_1bb}
&&1+\delta_M(\vec x,t)=\\ \nonumber
&&\quad \int d^3 q\; \delta^{(3)}(\vec x-\vec z(\vec q,t))\; \exp\left[\bar{\tilde f}_M(\d_iz^j(\vec q,\tin),\d_i\dot z^j(\vec q,\tin),\frac{\d}{\km},\epsilon(\vec z(\vec q,\tin),\tin),\dots,t)\right]\ .
\eea
This expression can indeed be thought as the symbolic writing of what the $N$-body codes solve numerically: the terms $\d_i z^j(\vec q,\tin)$ and $\d_i\dot z^j(\vec q,\tin)$ are related to nothing but the initial positions and velocities of the particles.
}.

Notice that, as explained in detail in~\cite{Porto:2013qua} in the Lagrangian-space EFTofLSS, defining correctly what is meant by the displacement field $\vec z(\vec q,t)$ in (\ref{eq:bias_lagrangian_1}) and (\ref{eq:delta_galaxy_1b}) at short distances requires the addition of counterterms that are supposed to correct for the uncontrolled contributions from short distances that appear when computing correlation functions of the displacement and of the matter density. Similar considerations apply for the terms such as $\d^2\phi(\vec z(\vec q,t))$ in  (\ref{eq:bias_lagrangian_1}) and (\ref{eq:delta_galaxy_1b})~\cite{Porto:2013qua}. In the same spirit as in~\cite{Senatore:2014via}, we are interested in performing the Lagrangian approach only for the purpose of resumming the contribution of the IR-displacements. For these IR effects, the counterterms of the Lagrangian EFT, which are important only in the UV, can be neglected for the purpose of discussion, and can be easily reinserted if necessary. As we will see shortly, we will rearrange the calculation so that this reinsertion is not necessary for practical purposes.

Let us proceed. If we define the displacement field as
\be
\vec s(\vec q,t)=\vec z(\vec q,t)-\vec q\ ,
\ee 
we can write (\ref{eq:bias_lagrangian_1}) in Fourier space for $\vec k\neq0$ as
\bea
&&\delta_M(\vec k,t)=\int d^3 q\; e^{- i\, \vec k\cdot\vec q}\\ \nonumber
&&\qquad \quad\times\;\exp\left[- i\, \vec k\cdot\vec s(\vec q,t)+f_M(\d_i\d_j\phi(\vec z(\vec q,\tin),\tin),\d_j v^i(\vec z(\vec q,\tin),\tin),\frac{\d}{\km},\epsilon(\vec z(\vec q,\tin),\tin),t)\right]\ .
\eea
This leads to the following expression for the power spectrum (similar expressions hold for the cross matter galaxies correlation)
\bea
&&\langle\delta_M(\vec k_1, t_1)\delta_M(\vec k_2, t_1)\rangle= (2\pi)^3 \delta^{(3)}(\vec k_1+\vec k_2)\\ \nonumber
&&\qquad \int d^3q\; e^{- i\vec k\cdot\vec q}\;\left\langle\exp\left[-i\, \vec k_1\cdot (\vec s(\vec q,t_1)-\vec s(\vec 0,t_2))\right.\right. \\ \nonumber
&&\quad\qquad\qquad\qquad\qquad+f_M(\d_i\d_j\phi(\vec q+\vec s(\vec q,\tin),\tin),\d_j v^i(\vec q+\vec s(\vec q,\tin),\tin),\frac{\d}{\km},\epsilon(\vec  q+\vec s(\vec q,\tin),\tin),t)\\ \nonumber
&&\quad\qquad\qquad\qquad\qquad\left.\left.+f_M(\d_i\d_j\phi(\vec s(\vec 0,\tin),\tin),\d_j v^i(\vec s(\vec 0,\tin),\tin),\frac{\d}{\km},\epsilon(\vec s(\vec 0,\tin),\tin),t)\right]\right\rangle\ .
\eea
We can now use the cumulant theorem to express the expectation value in terms of sum of connected $n$-point function in the following way
\bea
&&\langle\delta_M(\vec k_1, t_1)\delta_M(\vec k_2, t_1)\rangle= (2\pi)^3 \delta^{(3)}(\vec k_1+\vec k_2)\\ \nonumber
&&\qquad \int d^3q\; e^{- i\,\vec k\cdot\vec q}\;\exp\left[\left\langle\sum_{N=1}^\infty\frac{1}{N!}\left(-i\, \vec k_1\cdot (\vec s(\vec q,t_1)-\vec s(\vec 0,t_2))\right.\right. \right.\\ \nonumber
&&\quad\qquad\qquad\qquad\qquad+f_M(\d_i\d_j\phi(\vec q+\vec s(\vec q,\tin),\tin),\d_j v^i(\vec q+\vec s(\vec q,\tin),\tin),\frac{\d}{\km},\epsilon(\vec q+\vec s(\vec q,\tin),\tin),t)\\ \nonumber
&&\quad\qquad\qquad\qquad\qquad\left.\left.\left.+f_M(\d_i\d_j\phi(\vec s(\vec 0,\tin),\tin),\d_j v^i(\vec s(\vec 0,\tin),\tin),\frac{\d}{\km},\epsilon(\vec s(\vec 0,\tin),\tin),t)\right)^N\right\rangle_c\right]\ .
\eea
So far, this derivation of the Lagrangian bias is similar to what done in~\cite{Carlson:2012bu} in the case of a local Lagrangian bias. Here the derivation is different because we are taking a more general function $f_M$, as, at least in our opinion, this is necessary because of symmetries and renormalization, as we will explain later. More differences will appear subsequently.

Let us define $X_b$ as
\bea\label{eq:Xbdef}
&&X_b(\vec k,\vec q;t_1,t_2)=- \vec k_1\cdot (\vec s(\vec q,t_1)-\vec s(\vec 0,t_2))\\ \nonumber
&&\quad\qquad\qquad\qquad -i\, f_M(\d_i\d_j\phi(\vec q+\vec s(\vec q,\tin),\tin),\d_j v^i(\vec q+\vec s(\vec q,\tin),\tin),\frac{\d}{\km},\epsilon(\vec q+\vec s(\vec q,\tin),\tin),t)\\ \nonumber
&&\quad\qquad\qquad\qquad-i\,f_M(\d_i\d_j\phi(\vec s(\vec 0,\tin),\tin),\d_j v^i(\vec s(\vec 0,\tin),\tin),\frac{\d}{\km},\epsilon(\vec s(\vec 0,\tin),\tin),t)\ ,
\eea
and  $K_b$ as
\bea\label{eq:kbdef}
K_b(\vec k,\vec q;t_1,t_2)= \exp\left[\sum_{N=1}^{\infty} {i^N \over N! } \langle X_b(\vec k_1,\vec q;t_1,t_2)^N \rangle_c\right]\ ,
\eea
so that
\bea\label{eq:lagrangian_bias_2}
&&\langle\delta_M(\vec k_1, t_1)\delta_M(\vec k_2, t_1)\rangle= (2\pi)^3 \delta^{(3)}(\vec k_1+\vec k_2)\; \int d^3q\; e^{- i\vec k\cdot\vec q}\; K_b(\vec k,\vec q;t_1,t_2)\ .
\eea

Now we make the following manipulations, which are very similar at what we did in~\cite{Senatore:2014via} in the IR-resummed EFTofLSS, when we studied dark matter correlation functions, and which we follow even in the notation. The motivation is very similar. The reason why we wish to resum correlation functions is due to the displacements $\vec s$, or,  more precisely, the late time displacements, which in our universe are responsible for large effects that cannot be treated perturbatively. The functions $f_M$ are not functions of $\vec s$ without derivatives acting on it, unless when $\vec s$ is evaluated at very early times, where it is perturbatively small and can be Taylor expanded.  Of the terms in $K_b$, the leading terms that survive in the limit in which we send $\epsdm\to 0$ and $\epssM\to 0$ are the following
\bea\label{eq:kbdef}
K_{0}(\vec k,\vec q;t_1,t_2)= \exp\left[ -{1 \over 2 } \langle X_{0}(\vec k_1,\vec q;t_1,t_2)^2 \rangle\right]\ ,
\eea
where 
\be
X_{0}(\vec k,\vec q;t_1,t_2)=\vec k\cdot(\vec s(\vec q,t_1)_1-\vec s(\vec 0,t_2)_1)\ ,
\ee
is the function $X_b$ evaluated only with the linear displacements and neglecting completely the terms contained in $f_M$, as they indeed start with terms that vanish in the limit $\epsdm\to 0$ and $\epssM\to 0$.

We are interested in evaluating expression (\ref{eq:lagrangian_bias_2}) at all orders in the linear long-wavelength displacement fields $\epsilon_{s<}$, and to order $N$ in powers of $\epsilon_{\delta<}$ or $\epsilon_{s>}$. Since we are  going to resum neither in $\epsilon_{\delta<}$ nor in $\epsilon_{s>}$, we can treat these two parameters as the same, and let us denote them simply as $\epsilon_{\delta<}$. Let us denote an expression evaluated {\it up to} order $N$  in $\epsilon_{\delta<}$, and all orders in $\epsilon_{s<}$, by the following
\be
\left. K_b(\vec k,\vec q;t_1,t_2) \right|_N \ .
\ee 
Instead, let us denote the same expression evaluated {\it up to} order $N$ by expanding both in $\epsilon_{\delta<}$ and in $\epsilon_{s<}$, and counting them on equal footing, as
\be
\left. \left. K_b(\vec k,\vec q;t_1,t_2) \right|\right|_N \ .
\ee 
Since we are interested in resumming only the linear displacements, $X_{0}$ contains all the relevant information we wish to resum out of the exponential in~(\ref{eq:kbdef}). Once we have $K_{0}$, we can use it to do the following manipulations, valid up to order $N$ in $\epsilon_{\delta<}$. Since at all orders in $\epsilon_{s<}$ and leading order in $\epsilon_{\delta<}$, $K_b$ and $K_{0}$ are equal, we can  multiply and divide by $K_{0}$, and Taylor expand $K_b/K_{0}$ in powers of $\epsilon_{\delta<}$ and $\epsilon_{s<}$. In formulas, we have
\bea\nonumber\label{eq:kapprox}
&&\left. K_b(\vec k,\vec q;t_1,t_2) \right|_N \simeq K_{0}(\vec k,\vec q;t_1,t_2)\cdot\left.\left. {K_b(\vec k,\vec q;t_1,t_2)\over K_{0}(\vec k,\vec q;t_1,t_2)}\right|\right|_N=  \sum_{j=0}^N F_{||_{N-j}}(\vec k,\vec q;t_1,t_2)\cdot K_b(\vec k,\vec q;t_1,t_2)_{j} \ ,\\ 
\eea
where we defined
\be
F_{||_{N-j}}(\vec k,\vec q;t_1,t_2)=K_0(\vec k,\vec q;t_1,t_2) \cdot\left.\left.K_0^{-1}(\vec k,\vec q;t_1,t_2)\right|\right|_{N-j} \ .
\ee
Here the subscript ${}_j$ means that we take the order $j$ in $\epsilon_{\delta<}$ {\it and} $\epsilon_{s<}$ of a given expression, no to be confused with~${}||_j$, which  means instead that we take all terms of the same expression {\it up to} order~$j$ in~$\epsilon_{\delta<}$ {\it and} $\epsilon_{s<}$. We remind that expanding at a given overall order in $\epsilon_{\delta<}$ {\it and} $\epsilon_{s<}$ is nothing but the usual expansion in powers of the power spectrum, where we do not distinguish between factors of  $\epsilon_{\delta<}$ and of $\epsilon_{s<}$. This is the result obtained by performing the calculation in the Eulerian approach. We notice that the function $F_{||_{N-j}}(\vec k,\vec q;t_1,t_2)$ is exactly the same as the one that appears in~\cite{Senatore:2014via} for the resummation of the IR-effects in the dark matter correlation functions.

The final result of the above expression has the following useful property.  By construction, the two expressions agree up to order $N$ in $\epsilon_{\delta<}$, differing only for terms of  order higher than $N$ in $\epsilon_{\delta<}$, that we do not compute anyway. This is so because,  if we take $\epsilon_{s<}=0$, up to order $N$ in $\epsilon_{\delta<}$ the two expressions are identical by construction, as all terms up to order $N$ from $K_0$ have been designed to cancel identically. The approximate formula (\ref{eq:kapprox}) agrees with $\left. K_b(\vec k,\vec q;t_1,t_2) \right|_N$ to all order in $\epsilon_{s<}$ in the limit in which the long displacements are treated as free. This is another useful property of the above expression~\footnote{In~\cite{Senatore:2014via}, below eq.~(36), we described in detail why this the only term that needs a resummation. We also described how this resummation can be extended to include the displacements at the mildly non-linear level. Exactly the same discussion applies here.}. Equation (\ref{eq:kapprox}) is almost identical to (42) in~\cite{Senatore:2014via}, apart for the term $K_b$, which in~\cite{Senatore:2014via} is substituted by $K$. We can therefore make the same manipulations to arrive at the following equations. For the real space correlation functions, we have
\bea\label{eq:Lagrangian_bias_4} 
&&\left.\xi_{\delta_M\delta_M}(\vec r;t_1,t_2)\right|_N  =\sum_{j=0}^N \int d q\, q^2 \; P_{{\rm int }||_{N-j}}( r| q;t_1,t_2) \;  \xi_{\delta_M\delta_M,\,j}( q,t_1,t_2)\ .  
\eea
Here $\xi_{\delta_M\delta_M,\,j}$ is the power spectrum computed at order $j$ in the Eulerian approach. $P_{{\rm int}||_{N-j}}$ is defined as
\be
P_{{\rm int}||_{N-j}}(r|q;t_1,t_2)=2\pi \int_{-1}^{1} d\mu\; \int\frac{d^3 k}{(2\pi)^3}\; e^{-i \vec k\cdot (\vec q-\vec r)}\;F_{||_{N-j}}(\vec q,\vec k;t_1,t_2)\ ,
\ee
where $\mu$ is the cosine of the angle between $\vec q$ and $\vec r$. $P_{{\rm int}||_{N-j}}(r|q;t_1,t_2)$ can be thought of as the probability of starting at Lagrangian distance $q$ and ending up at physical distance $r$. It is exactly the same function that appears in the resummation of the dark matter correlation functions. Similarly, in Fourier space, we have
\be\label{eq:Lagrangian_bias_5} 
\left.P_{\delta_M\delta_M}(k;t_1,t_2)\right|_N=\sum_{j=0}^N \int \frac{d^3k'}{(2\pi)^3}\; M_{||_{N-j}}( k, k';t_1,t_2)\; P_{\delta_M\delta_M,\,j}(k';t_1,t_2)\ .
\ee
where $M_{||_{N-j}}( k, k';t_1,t_2)$ is defined as the double Fourier transform of $ P_{{\rm int}||_{N-j}}(r|q;t_1,t_2)$
\bea
&&M_{||_{N-j}}( k, k';t_1,t_2)=\frac{1}{4\pi}\int d^3 r\; d^3 q \; P_{{\rm int}||_{N-j}}(r|q;t_1,t_2) \; e^{i \vec k \cdot\vec r} \;e^{-i \vec k' \cdot\vec q}\\ \nonumber
&&\qquad\qquad\qquad \ \qquad=\int d^3 q \; F_{||_{N-j}}(\vec q,-\vec k;t_1,t_2) \; e^{i (\vec k-\vec k') \cdot\vec q}\ ,
\eea
and $P_{\delta_M\delta_M,\,j}$ is the $j$-th order term for the power spectrum in the Eulerian calculation. The simplification in the second line, that reduces the computation from a bi-dimensional Fourier Transform to performing a one-dimensional Fourier transform for each $k$ of interest, has appeared in~\cite{Angulo:2014tfa}.

In summary, we have been able to write the IR-resummed power spectrum of collapsed objects as a convolution of the several terms that one obtains in the Eulerian calculation. Similarly, the IR-resummed correlation function in real space is given by a convolution of the terms that appear in the correlation function when computed in the Eulerian approach.

Exactly the same formula as (\ref{eq:Lagrangian_bias_5}) can be obtained if we had started from a different formulation of the Lagrangian bias, in which, instead of starting from (\ref{eq:delta_galaxy_1b}), we write the galaxy overdensity in terms of the late time matter fields, in a non-local in time way:
\bea\label{eq:delta_galaxy_2}
&&1+\delta_M(\vec x,t)= \int d^3 q\; \delta^{(3)}(\vec x-\vec z(\vec q,t))\;\\ \nonumber
&& \qquad\quad\times\  \exp\left[\tilde f_M\left(\int^t d t'\;K_{\d_i\d_j\phi}(t,t')\;\d_i\d_j\phi(\vec z(\vec q,t'),t'),\right.\right. \\ \nonumber
&&\qquad\qquad\qquad\qquad\ \left.\left.\int^t dt'\;K_{\d_iv^j}(t,t')\;\d_j v^i(\vec z(\vec q,t'),t'),\frac{\d_\xfl}{\km},\int^t dt'\;K_{\epsilon}(t,t')\;\epsilon(\vec z(\vec q,t'),t'),t\right)\right]\ ,
\eea
where $K_{\ldots}(t,t')$ are some kernels.
This formula tells us that the galaxy overdensity at a given location $\vec x$ is proportional to the tidal tensors, velocity gradients, etc, evaluated on the past trajectory of the dark matter fluid element that ended up at the considered location $\vec x$. It is clear that the difference between this formula and the former one is just a potentially complicated re-definition of $f_M$.
We can then proceed with the same derivation and approximations that we did to derive (\ref{eq:Lagrangian_bias_5}) by noticing the following. The reason why we need to resum the displacement fields is because we need to be sure to evaluate the displacement at the relevant~$\vec q$ than ended up at $\vec x$, and also the fields inside $\tilde f_M$ at the right location $\vec x$ as a function of $\vec q$. The approximations we did to derive (\ref{eq:Lagrangian_bias_5}) amount to approximately evaluating the $\delta^{(3)}(\vec x-\vec z(\vec q,t))$ as $\delta^{(3)}(\vec x-q-\vec s(\vec q,t)_1)$, and then recovering perturbatively the right $\delta$-function by Taylor expanding in $\tilde\epsilon_{s<}\sim (k|\vec s-\vec s_1|)^2$ further than in~$\epsilon_{\delta<}$ and $\epsilon_{s>}$. This means that exactly the same approximations in the derivation that lead to~(\ref{eq:Lagrangian_bias_5}) are applicable if we had started from (\ref{eq:delta_galaxy_2}), leading, indeed, to~(\ref{eq:Lagrangian_bias_5}) in an unchanged way.

Yet another way to obtain the same formula (\ref{eq:Lagrangian_bias_5}) is the following. In the Lagrangian-space EFTofLSS, dark matter regions are described as extended particles labelled by $\vec q$ and endowed with multiples: $Q^{ij}(\vec q)$, $Q^{ijk}(\vec q),\,\ldots$. They induce overdensities of dark matter not only through their clustering,  but also through the energy associated to the deformations of their multipoles~\cite{Porto:2013qua}
\beq
\label{equation2lagrang}
 \frac{3}{2} {\cal H}^2 \Omega_m \delta_{m}({\vec x},t)\equiv \d^2\Phi(\vec x,t) = \frac{3}{2} {\cal H}^2 \Omega_m\left(\delta_{n}({\vec x},t)  + \frac{1}{2} \partial_i\partial_j  {\cal Q}^{ij}({\vec x},t) 
- \frac{1}{6} \partial_i\partial_j \partial_k {\cal Q}^{ijk}({\vec x},t) + \cdots\right)\ ,
\eeq
where the matter density $\delta_{m}$ is to be distinguished from the number density $\delta_{n}$, and where real-space multipole moments are given by:
\bea
\label{smooth1lagrang}
1+\delta_{n}({\vec x},t)&\equiv& \int  d^3{\vec q}\;\delta^3( {\vec x} - {\vec z}({\vec q},t))\ , \nn \\
\label{calq}  {\cal Q}^{i_1 \dots i_p}({\vec x},t) &\equiv& \int d^3 {\vec q}\; Q^{i_1\ldots i_p}({\vec q},t)\;\delta^3({\vec x}-{\vec z}({\vec q},t))\ .
\eea 
Since the gravitation field at a given location is proportional also to the deformation of the multiples of the extended particles that ended up at a given location, and since it is exactly this kind of terms that induce the collapse of objects, we can write an analogous formula to~(\ref{eq:delta_galaxy_1b}) as
\bea\label{eq:delta_galaxy_1c}
&&1+\delta_M(\vec x,t)=\int d^3 q\; \delta^{(3)}(\vec x-\vec z(\vec q,t))\;\\ \nonumber
&&  \qquad\qquad \times \exp\left[\bar f_M\left(\int^t d t'\;K_{\d_j s^i}(t,t')\, \d_j s^i(q,t'), 
\ldots,\right.\right.\\ \nonumber 
&&\quad\quad \qquad\qquad  \qquad\left.\left. \int^t d t'\;K_{\d_{i}\d_{j}Q^{ij}}(t,t')\,\d_{i}\d_{j}Q^{ij}(\vec q,t '), \int^t d t'\;K_{\d_{i}\d_{j}\d_{l}Q^{ijl}}(t,t')\,\d_{i}\d_{j}\d_{l}Q^{ijl}(\vec q,t' ),\ldots, \right.\right.\\\nonumber
&&\quad\quad \qquad\qquad  \qquad\left.\left.\int^t d t'\;K_{\epsilon}(t,t')\,\epsilon(\vec q,t'),\ldots,t\right)\right]\ .
\eea
In this way of writing the Lagrangian bias, at the cost of introducing the notation of the Lagrangian-space EFTofLSS, it has become even more immediate that one can perform all the same approximations as we did just above, leading to~(\ref{eq:Lagrangian_bias_5}), in an unchanged way. These are all different but equivalent definitions of the Lagrangian bias that coincide once we express the result in terms of a resummation of the Eulerian calculation, as in~(\ref{eq:Lagrangian_bias_5}).

There is one further subtlety to complete the relevant equations. As explained in detail in~\cite{Senatore:2014via}, a discussion that applies here practically unaltered, when computing correlation functions involving fields that are not scalars under time dependent space diffeomorphisms, such as the momentum correlation function, some of the Eulerian terms in~(\ref{eq:Lagrangian_bias_4}) or~(\ref{eq:Lagrangian_bias_5}) that in the naive Eulerian counting are considered higher orders, should instead be included. These are the terms that are manifestly IR-sensitive. For example, in the analogous formula for the power spectra of the momentum divergence of the collapsed objects there is a term proportional to $\d_i\pi^i_M\supset d_{v} \d_i\delta_M \,v^i\sim d_v c_{\d^2\phi}\, \d_i\delta\, v^i$ which is IR-sensitive. In that case, we need to upgrade our IR-resummed formulas to
\bea
\label{eq:momentum3}
&&\left.P_{\pi_M\,\pi_M}(k;t_1,t_2)\right|_N=\\ \nonumber
&& \int \frac{d^3k'}{(2\pi)^3}\; \left[\sum_{j=0}^N M_{||_{N-j}}( k, k';t_1,t_2)\;  P_{\pi_M\,\pi_M,\,j}(k';t_1,t_2)+ M_{||_{0}}( k, k';t_1,t_2)\;  \Delta P_{\pi_M\,\pi_M,\,N}(k';t_1,t_2)\right]\ .
\eea
where $\Delta P_{\pi\,\pi,\,N}(k;t_1,t_2)$ corresponds to adding all terms of order $N$ in $\epsilon_{\delta<}$ that were not included in the standard Eulerian calculation because we considered $v^i$ as a perturbation of order $(\epsilon_{\delta<})^{1/2}$. 
For example, at leading order, in the equal-time momentum-divergence power spectrum, we need to add only one term, which is given by
\bea \label{eq:momentum_final}
&&\Delta P_{\pi_M\,\pi_M,\,N}(k;t_1,t_1)= d_v^2 c_{\d^2\phi}^2 \frac{1}{a^2} \left[\langle v(\vec x_1,t_1)^i v(\vec x_2, t_1)^j \rangle_{1} \; \langle\d_i \delta(\vec x_1,t_1)\d_j \delta(\vec x_2, t_1) \rangle_{N}\right]_k\\  \nn
&&\qquad\qquad= d_v^2 c_{\d^2\phi}^2\frac{1}{a^2}\int^{\bar\Lambda_{\rm Resum}(k)} \frac{d^3 k'}{(2\pi)^3} \;\; P_{\theta\theta,\, 1}(\vec k',t_1,t_1)\; \frac{\left(\vec k'\cdot (\vec  k-\vec k')\right)^2}{k'^4} \;P_{\delta\delta,\,N+1}(|\vec k-\vec k'|,t_1,t_1)\ ,
\eea
where for simplicity we have approximated the kernels as local in time.

By applying these Lagrangian, or IR-resummed, formulas, no expansion in $\epsilon_{s<}$ is present, and the perturbative expansion is therefore an expansion is just in powers of $k/\knl$ and $k/\km$, which makes the perturbative scheme manifestly convergent to the true answer for $k\ll\knl$ and $k\ll \km$~\footnote{As described in~\cite{Senatore:2014via}, since the IR-resummation is not performed exactly, a small residual effect from the long modes is included order by order in perturbation theory, which corresponds to having an additional expansion parameter. In the notation of~\cite{Senatore:2014via}, this parameter is called $\tilde\epsilon_{s<}$, and it is much smaller than $\epsilon_{s<}$.  In practice, the parameter $\tilde\epsilon_{s<}$ is so small that it can be neglected in the counting. Finally, by convergence to the true answer we mean it in the usual sense of asymptotic series, that is up to when non-perturbative effects become important, something that is expected to happen very close to $\knl$ or~$\km$.}.

\section{Perturbative Expansion for Correlation Functions}

In this section  we wish to explain how the contribution of the various terms can be classified in powers of $k/\knl$ and $k/\km$. If we perform the calculation in the Lagrangian formalism, the expansion parameter $\epssm$ does not appear. In the former section, we showed how the Lagrangian calculation  can actually be implemented by re-adapting the Eulerian calculation. Because of this, even though in the Eulerian calculation the parameter $\epssm$ does appear, we can ignore it, assuming we will ultimately actually perform the Lagrangian calculation. In~\cite{Carrasco:2013mua} (see also~\cite{Pajer:2013jj}), it has been shown that the universe can be well approximated by a piecewise scaling one, with power spectrum
\bea\label{eq:fit}
P_{11}(k)=(2\pi)^3 \left\{ \begin{array}{ll} 
\frac{1}{\knl^3}\left(\frac{k}{\knl}\right)^{-2.1} & \text{ for} \ {k>\ktr} \ , \\
 \frac{1}{\tknl^3}\left(\frac{k}{\tknl}\right)^{-1.7}   & \text{for} \ {k<\ktr} \ ,
\end{array} \right.
\eea
where $\tknl = (\knl^{0.9} \ktr^{0.4})^{1/1.3}$ and $\ktr$ is the transition scale between the two different power-law behaviors. The fit parameters are given by
\beq
\knl = 4.6 \invMpc\ ,  \qquad \ktr = 0.25 \invMpc \ ,\qquad \tknl = 1.8 \invMpc \ .
\eeq

Depending on the $k$'s of interest, the result of correlation functions of dark matter scales as we perform higher order corrections as powers of $k/\knl$. For example, for the power spectrum we have:
\bea\label{eq:dark_matter_expansion}
&&\langle\delta(\vec k)\delta(\vec k)\rangle'\sim\\ \nonumber
&&\qquad \langle\delta(\vec k)\delta(\vec k))\rangle'_{\rm tree} \times\underbrace{\left[1+ \left(\frac{k}{\knl}\right)^2+\ldots+\left(\frac{k}{\knl}\right)^D \right]}_\text{Derivative Expansion}\underbrace{\left[1+\left(\frac{k}{\knl}\right)^{(3+n)}+\ldots+\left(\frac{k}{\knl}\right)^{(3+n)L}\right]}_\text{Loop Expansion}\\ \nonumber
&&\qquad+\underbrace{ \left[\left(\frac{k}{\knl}\right)^{4}+\left(\frac{k}{\knl}\right)^{6}+\ldots\right]}_\text{Stochastic Terms}\ ,
\eea
where we have assumed that $k>\ktr$, as otherwise one should simply replace $\knl$ with $\tknl$ in the loop expansion (that is, it should not be replaced in the derivative expansion or in the stochastic terms). The $\langle\rangle'$ means that we have removed a momentum-conservation $\delta$-function from the result of the expectation value.  Here we have neglected the numerical order one coefficients. The first squared parenthesis represents the contribution from the counterterm associated to the response of the short distance dark matter stress tensor to long wavelength fluctuations, evaluated at higher and higher order in derivatives up to order $D$~\footnote{Or from the stochastic terms when used in some non-linear diagram, something that happens at very high order in perturbation theory.}. The second squared parenthesis represents the contribution from performing higher loops, that contribute as powers of $k^3P(k)\sim (k/\knl)^{3+n}$, where $n$ is the slope of the approximately scaling power spectrum: $n\simeq -1.7$ for $k\lesssim \ktr$, $n\simeq -2.1$ for $k\gtrsim \ktr$. The last term represents the contribution of the stochastic counterterm. Because of matter and momentum conservation, the stochastic terms start with $(k/\knl)^4$. Here, for clarity and synthesis, we have neglected all time-dependent coefficient themselves: different terms in perturbation theory carry different time dependent factors.  In practice, one can think of the above expression as evaluated at redshift zero. It is straightforward to extend it to different redshift. Similar considerations apply for the remaining equations in this section.

There is an important point to understand how formula (\ref{eq:dark_matter_expansion}) emerges when we do the actual calculation. As explained in detail in the papers that established the EFTofLSS as applied to dark matter calculations~\cite{Carrasco:2012cv,Pajer:2013jj,Carrasco:2013mua},  perturbation theory reorganizes itself in an expansion in powers of $k/\knl$ only {\it after} renormalization, that is only after the UV-sensitive terms, i.e. the terms that in a scaling universe would be divergent as powers of the UV cutoff $\Lambda$, have been reabsorbed into the counterterms, and the counterterms have been chosen to agree with the data at some observation point.

If we now pass to the collapsed objects, we will see that on top of the parameter $k/\knl$, we have also the parameter $k/\km$.  In particular, let us compute the galaxy dark-matter equal-time power spectrum. For the propose of estimating, we can neglect all numeral coefficients and in particular the non-locality in time. We schematically have
\bea
&&\langle\delta_M(\vec k)\delta(\vec k)\rangle'\\ \nonumber
&&\quad
\sim c_{\d^2\phi} \langle\delta(\vec k)\delta(\vec k)\rangle'+c_{\d^2\phi\d^2\phi}\langle[\delta^2]_{\vec k} \delta(\vec k)\rangle'+\, c_{\d^4\phi}  \left(\frac{k}{\km}\right)^2 \langle\delta(\vec k)\delta(\vec k)\rangle' + \frac{1}{\knl^{3/2} \km^{3/2}}\frac{k^2}{\knl^2}+\ldots\ ,  
\eea
where the first terms comes from the contraction of the linear bias term, the second from the contraction of the non-linear bias term, the third from the higher derivative terms, and finally the fourth from the stochastic terms. $[\delta^2]_{\vec k}$ means that we take the $\vec k$ component of the product field $\delta(\vec x,t)^2$. The last term represents the contribution from the stochastic term. 
To clarify the perturbative expansion, we can schematically rearrange the terms in the following way
\bea\label{eq:dark_matter_galaxy_expansion}
&&\langle\delta_M(\vec k)\delta(\vec k)\rangle'\sim P_{11}(k_1)  \\ \nonumber
&&\quad \times\left\{\underbrace{\left[c_{\d^2\phi}+c_{\d^4\phi}\left(\frac{k}{\km}\right)^2+\ldots+c_{\d^{2D}\phi}\left(\frac{k}{\km}\right)^{2D-2}\right]}_\text{Linear Bias Derivative Expansion}\underbrace{\left[1+\left(\frac{k}{\knl}\right)^{(3+n)}+\ldots+\left(\frac{k}{\knl}\right)^{(3+n)L}\right]}_\text{Matter Loop Expansion}\right.
\eea
\bea \nonumber
&&\qquad+\underbrace{\left[c_{(\d^2\phi)^2}+c_{\d^2(\d^2\phi)^2}\left(\frac{k}{\km}\right)^2+\ldots+c_{\d^{2D-2}(\d^2\phi)^2}\left(\frac{k}{\km}\right)^{2D-2}\right]}_\text{Quadratic Bias Derivative Expansion}  
\\ \nonumber
&&\quad\qquad\qquad\qquad\qquad\qquad\times\left.\underbrace{\left(\frac{k}{\knl}\right)^{3+n}}_\text{Quadratic Bias}\underbrace{\left[1+\left(\frac{k}{\knl}\right)^{(3+n)}+\ldots+\left(\frac{k}{\knl}\right)^{(3+n)L}\right]}_\text{Matter Loop Expansion}\right. 
\\ \nonumber
&&\qquad+\underbrace{\left[c_{(\d^2\phi)^3}+c_{\d^2(\d^2\phi)^3}\left(\frac{k}{\km}\right)^2+\ldots+c_{\d^{2D-2}(\d^2\phi)^3}\left(\frac{k}{\km}\right)^{2D-2}\right]}_\text{Cubic Bias Derivative Expansion} \\ \nonumber 
&&\quad\qquad\qquad\qquad\qquad\qquad\left.\times\underbrace{ \left(\frac{k}{\knl}\right)^{2(3+n)}}_\text{Cubic Bias}\underbrace{\left[1+\left(\frac{k}{\knl}\right)^{(3+n)}+\ldots+\left(\frac{k}{\knl}\right)^{(3+n)L}\right]}_\text{Matter Loop Expansion}\right\}\\ \nonumber 
&&\qquad\left.+\underbrace{\left[c_{\epsilon_0}\, c_{m,\rm stoch, 1}+c_{a,\,\epsilon_2}\left(\frac{k}{\km}\right)^2+c_{m,\rm stoch, 2}\left(\frac{k}{\knl}\right)^2+\ldots\right]}_\text{Stochastic Bias Derivative Expansion}\; \underbrace{\frac{1}{(\km^3\knl^3)^{1/2}}\left(\frac{k}{\knl}\right)^2}_\text{Stochastic Bias}+\dots \right. \ .
\eea
This expression (\ref{eq:dark_matter_galaxy_expansion}) is similar to the one obtained for dark matter (\ref{eq:dark_matter_expansion}), and the same replacement $\knl\to\tknl$ should be meant in this case as well as $k\lesssim \ktr$ for all the expansions associated with the power spectrum (i.e. the higher derivative matter counterterms are always suppressed by $\knl$). Thanks to the Lagrangian treatment, the expansion for dark matter is just in powers of $k/\knl$, both for what concerns the non-linearites and the higher derivative terms. Instead, in the case of galaxies, the expansion in derivatives is controlled by the parameter $\km$. For very massive objects, we expect $\km\lesssim \knl$, while for relatively light objects, $\km\gtrsim \knl$. This suggest that for very light object the derivative expansion in the bias parameters is more suppressed than in dark matter, while for massive objects it is the opposite. 

Notice the last term that comes from the correlation of the stochastic bias for collapsed objects with the stochastic counterterm $\Delta\tau$ for dark matter: clearly, if there happens to be more dark matter in one given realization, it is expectable there will be more collapsed objects. The stochastic term for dark matter is constrained by matter and momentum conservation to appear in the equations of motion with two derivatives in front: $\d^2\Delta\tau$. This explains the factor of $k^2$. Since we have
\be
\langle\epsilon_k\epsilon_k\rangle\sim \frac{1}{\km^3}\left(1+\frac{k^2}{\km^2}+\ldots\right)\,, \qquad \langle[\d^2\Delta\tau]_k[\d^2\Delta\tau]_k\rangle\sim \frac{1}{\knl^3}\left(\frac{k}{\knl}\right)^4\left(1+\frac{k^2}{\knl^2}+\ldots\right)\ , 
\ee
we have taken
\be
\langle\epsilon_k[\d^2\Delta\tau]_k\rangle\sim \frac{1}{(\km^3\knl^3)^{1/2}}\left(1+\frac{k^2}{\km^2}+\frac{k^2}{\knl^2}+\ldots\right)\ . 
\ee
There are two points to comment about this behavior. First, the stochastic contribution starts with $k^0$ for collapsed objects. This is the familiar shot noise term. This behavior is different than the contribution of the stochastic term in dark matter, that starts with a much higher power of $k$, $k^4$, which makes it quite negligible in the dark matter correlation functions. The reason why the stochastic term in dark matter start with $k^4$ can be traced to matter and momentum conservation. Clearly, number and momentum are not conserved for collapsed objects, which is why the stochastic term is allowed to start at $k^0$. The other point we wish to comment on the stochastic term is the size of order $1/\km^3$. With this size, the shot noise is expected to induce a variance of order one on length scales of order $1/\km$, as indeed expected.

A similar expression holds for the power spectrum of compact object, with trivial differences. If we correlate objects of type $a$ and type $b$, we can write
\bea
&&\langle\delta_{M_a}(\vec k)\delta_{M_b}(\vec k)\rangle\sim \\ \nonumber
&&\quad(c_{a,\,\d^2\phi})c_{b,\,\d^2\phi}) \langle\delta(\vec k)\delta(\vec k)\rangle+c_{a,\,(\d^2\phi)^2} c_{b,\,\d^2\phi}\langle[\delta^2]_{\vec k} \delta(\vec k)\rangle+\, c_{a,\,\d^2\phi}c_{b,\,\d^4\phi}  \left(\frac{k}{\km}\right)^2  \langle\delta(\vec k)\delta(\vec k)\rangle+\ldots  ,
\eea
which leads to the following power series
\bea\label{eq:dark_galaxy_galaxy_expansion}
&&\langle\delta_M(\vec k)\delta_M(\vec k)\rangle'\sim P_{11}(k_1)  \\ \nonumber
&&\quad \times\left\{\underbrace{\left[c_{a,\,\d^2\phi} c_{b,\,\d^2\phi}+c_{a,\,\d^2\phi}c_{b,\,\d^4\phi}\left(\frac{k}{\km}\right)^2+\ldots\right]}_\text{Linear-Linear Bias Derivative Expansion}\times \underbrace{\left[1+\left(\frac{k}{\knl}\right)^{(3+n)}+\ldots\right]}_\text{Matter Loop Expansion}\right.
\\ \nonumber
&&\qquad+\underbrace{\left[c_{a,\,(\d^2\phi)^2}c_{b,\,\d^2\phi}+c_{a,\,\d^2(\d^2\phi)^2}c_{b,\,\d^2\phi}\left(\frac{k}{\km}\right)^2+\ldots\right]}_\text{Linear-Quadratic Bias Derivative Expansion} \\ \nonumber 
&&\qquad\qquad\qquad\qquad\qquad\qquad\times\left.\underbrace{\left(\frac{k}{\knl}\right)^{3+n}}_\text{Quadratic Bias}\underbrace{\left[1+\left(\frac{k}{\knl}\right)^{(3+n)}+\ldots+\left(\frac{k}{\knl}\right)^{(3+n)L}\right]}_\text{Matter Loop Expansion}\right. 
\\ \nonumber
&&\qquad+\underbrace{\left[c_{a,\,(\d^2\phi)^3}c_{b,\,\d^2\phi}+c_{a\,,(\d^2\phi)^2}c_{b,\,(\d^2\phi)^2}+\left(c_{a,\,\d^2(\d^2\phi)^3}c_{b,\,(\d^2\phi)^3}+c_{a,\,\d^2(\d^2\phi)^2}c_{b,\,(\d^2\phi)^2}\right)\left(\frac{k}{\km}\right)^2\ldots\right]}_\text{Linear-Cubic \& Quadratic-Quadratic Bias Derivative Expansion} \\ \nonumber 
&&\left.\qquad\qquad\qquad\qquad\qquad\qquad\times\underbrace{ \left(\frac{k}{\knl}\right)^{2(3+n)}}_\text{Cubic Bias}\underbrace{\left[1+\left(\frac{k}{\knl}\right)^{(3+n)}+\ldots+\left(\frac{k}{\knl}\right)^{(3+n)L}\right]}_\text{Matter Loop Expansion}\right\} 
\\ \nonumber 
&&\qquad+\underbrace{\left[c_{a,\,\epsilon_0}c_{b,\,\epsilon_0}+c_{a,\,\epsilon_2}c_{b,\,\epsilon_0}\left(\frac{k}{\km}\right)^2+\ldots\right]}_\text{Stochastic Bias Derivative Expansion} \underbrace{\frac{1}{\km^3}}_\text{Stochastic Bias} \\ \nonumber
&&\qquad+\underbrace{\left[c_{a,\,\epsilon\,\d^2\phi}c_{b,\,\epsilon\,\d^2\phi}+c_{a,\,\epsilon\,\d^2\d^2\phi}c_{b,\,\epsilon\,\d^2\phi}\left(\frac{k}{\km}\right)^2\ldots\right]}_\text{Linear-Stochastic Bias Derivative Expansion}   \underbrace{\frac{1}{\km^3}}_\text{Stochastic Bias}\\ \nonumber
&&\qquad\qquad\qquad\qquad\qquad\qquad\times\left.\underbrace{\left(\frac{k}{\knl}\right)^{3+n}}_\text{Quadratic Bias}\underbrace{\left[1+\left(\frac{k}{\knl}\right)^{(3+n)}+\ldots+\left(\frac{k}{\knl}\right)^{(3+n)L}\right]}_\text{Matter Loop Expansion} +\dots \right. \ ,
\eea
where for simplicity we did not explicitly distinguish between $k_{M_a}$ and $k_{M_b}$, which are in general different.
This expression is very similar to the one for the correlation functions, except for being quadratic in the bias coefficients.

Another aspect of the above formulas we want to explicitly comment is the following point related to the appearance of the dark matter power spectrum in the correlation functions of the collapsed objects. We have expressed the non-linear corrections due to the dark matter non-linearities simply as $P_{\delta\delta}(1+(k/\knl)^{3+n}+\ldots)$. It should be made clear that this is a schematic representation which is correct just for estimating the size of the various contributions in powers of $k/\knl\ll1$, $k/\km\ll 1$. However, it is not true that the dark matter non-linearities combine themselves so that the correlation functions of collapsed objects contains terms just proportional to the non-linear dark matter power spectrum. In particular, for example focussing on the cross-correlation, one could imagine that the term linear in the bias should give $P_{\delta\delta, \rm NL}$, which represents the non-linear dark matter power spectrum. This is not correct for the following reason. As we stressed in~(\ref{eq:euler_bias_2}), the bias coefficient are to be interpreted as non-local time kernels. Therefore, what contributes to the cross correlation is rather something of the form
\bea\label{eq:help_1}
\langle\delta_M( k,t)\delta( k,t)\rangle'\simeq \int^t dt' \; H(t')\left[   \bar c_{\d^2\phi}(t,t')\; \frac{1}{H(t')^2}\langle\delta(k,t')\delta(k,t')\rangle' \right]\ .
\eea
When we evaluate $\langle\delta(k,t)\delta(k,t)\rangle'$ in perturbation theory, we have
\be\label{eq:help_2}
\langle\delta_M(k,t)\delta(k,t)\rangle'\simeq\frac{D(t)^2}{D_0{}^2}\langle\delta(k,t_0)\delta(k,t_0)\rangle'_\text{tree}+\frac{D(t)^4}{D_0{}^4}\langle\delta(k,t_0)\delta(k,t_0)\rangle'_\text{1-loop}+\ldots\ ,
\ee
where $D$ is the growth factor, and the subscript $_0$ means that the quantity is evaluated at present time. When we substitute~(\ref{eq:help_2}) into~(\ref{eq:help_1}), we realize that the time integrals can be formally done, to obtain an expression of the following form
\be
\langle\delta_M(\vec k,t)\delta(k,t)\rangle'\simeq c_{\d^2\phi,1}(t) \; \langle\delta(k,t)\delta(k,t)\rangle'_\text{tree}+ c_{\d^2\phi,2}(t)\; \langle\delta(k,t)\delta(k,t)\rangle'_\text{1-loop}+\ldots\ ,
\ee
where
\bea
&&c_{\d^2\phi,1}(t)=\int^t dt' \; H(t')\left[   \bar c_{\d^2\phi}(t,t')\; \frac{1}{H(t')^2} \frac{D^2(t')}{D^2(t)}\right]\ ,\\ \nonumber
&&   c_{\d^2\phi,2}(t)=\int^t dt'\; H(t') \left[   \bar c_{\d^2\phi}(t,t')\; \frac{1}{H(t')^2} \frac{D^4(t')}{D^4(t)}\right]\ , \qquad \ldots\ .
\eea
It is clear that in general $c_{\d^2\phi,1}, \, c_{\d^2\phi,2},\,\ldots$ are comparable, but non-equal, numbers. Therefore the matter contribution does not recombine into the non-linear power spectrum.  Similar treatment trivially extends to the other correlation functions. 

This observation is also useful because it tells us how we can effectively treat the non-locality in time of the bias coefficients. Since the time-dependence in perturbation theory is $k$-independent, non-locality in time amounts to simply consider the various terms in perturbation theory as multiplied by independent coefficients. Very explicitly, if the theory was local in time, various terms that are related to one another by  representing the contribution of a given order in perturbation theory to the same quantity, would be multiplied by the same coefficient. Instead, when the theory is non-local in time, these same various terms are multiplied by a different numerical coefficient.  This gives an efficient treatment of the non-locality in time, which is identical to the one that was developed for the correlation functions of dark matter in~\cite{Carrasco:2013mua}. Since the time scales involved in the problem are of order Hubble, it is also possible to effectively parametrize the kernels with a small number of free coefficients, so that the number of free parameters can be reduced~\cite{Carrasco:2013mua}. As we argued in sec.~\ref{sec:quasi-local}, there is an approximate, not exact, sense in which the theory can be treated as local in time. In this case, higher time-derivative terms need to appear in the description of the theory and they are multiplied by new coupling constants.
 
 Finally, this expression teaches us how to measure the bias parameters from observations or $N$-body simulations. In the case of the local-in-time, there is one unique parameter for a given dark matter correlation function. Instead, in the non-local-in-time treatment, there is one parameter for each different term in the perturbative expansion of each dark matter correlation function that appears in the expressions. In practice, there is no need to measure the kernels such as $ \bar c_{\d^2\phi}(t,t')$, but it is enough to measure $c_{\d^2\phi,1}, \, c_{\d^2\phi,2},\,\ldots$. A measure of the non-locality in the theory is indeed to see how different are $c_{\d^2\phi,1}$ and $c_{\d^2\phi,2} $.

 \section{Dark-Matter Galaxies Cross Correlation\label{sec:dark-matter-cross}}

In the former section, we explained how to estimate the size for he various terms that appear in the perturbative expansion of the EFTofLSS when applied to biased objects.  It is interesting to give more explicitly the relevant expressions for the correlation function between dark-matter and collapsed objects, and for the power spectrum of collapsed objects, and to determine the actual bias parameters that are necessary at a given order.
 
 For the correlation functions of dark matter, the EFT calculation carried out up to to two-loops with IR-resummation matches to $1\,\%$ accuracy the observations from $N$-body simulations up to $k\simeq 0.6\invMpc$ with only one coupling constant, $c_s$~\cite{Senatore:2014via,Carrasco:2013mua}. At one-loop, the prediction stops being $1\,\%$ accurate at about $k\simeq 0.3\invMpc$.

Since the two-loop matter power spectrum has been already computed, we could in principle include everything whose contribution is smaller than $1\%$ up to $k\simeq 0.6\hinvMpc$.  However, for the purpose of this paper, we will stop at one loop, and therefore we will include everything that is smaller then $1\%$ at $k\simeq 0.3\invMpc$.

Before proceeding, there is an important observation to do. When computing quantities in perturbation theory, as explained in detail in~\cite{McDonald:2009dh}, one can usefully use the fact that up to a given order in perturbation theory some different quantities turn out to be numerically equal in order to simplify the algebra. Since in this paper we will stop at  one loop, it is sufficient for us to use all the relationships that relate quantities up to third order. 

Up to linear order, we have that $\d_i v^i=-\frac{D'}{D} \delta$, where $\H=a H$, and $'=\d/\d\tau$, with $\tau$ being the conformal time. As usual in SPT, we can redefine the velocity field as 
$\theta\equiv\d_i\tilde v^i=-\frac{D}{D'} \d_i v^i$, so that $\theta=\delta$ at linear level. For simplicity, we also redefine $\phi$ so that $\d^2\phi=\delta$. So, instead of using $v^i$ as an independent variable, we can replace $\theta$ with 
\be
\eta(\vec x,t)=\theta(\vec x,t)- \delta(\vec x,t)
\ee
and $\d_j v^i$ with $t_{ij}$ defined as
\be
t_{ij}(\vec x,t)=\d_i v_j(\vec x,t)-\frac{1}{3}\delta_{ij} \theta(\vec x,t) -s_{ij}(\vec x,t)\ .
\ee
$\eta$ and $t_{ij}$ start at second order in perturbation theory. Notice that since vorticity is generated only at very high order in perturbation theory~\cite{Carrasco:2013mua} (see also~\cite{Mercolli:2013bsa}), $t_{ij}$ can be considered symmetric in $i$ and $j$ up to a very high order in perturbation theory. Similarly, we replace $\d^2\phi=\delta$ with $\delta$. Let us define also the traceless tidal tensor as 
\be
s_{ij}=\d_i\d_j\phi-\frac{1}{3} \delta_{ij}\, \delta \ .
\ee
Up to second order in perturbation theory, we have that $\eta_{(2)}=\frac{2}{7}s^{(1)}{}^2-\frac{4}{21}\delta^{(1)}{}^2$, so that we can define a quantity $\psi$ that is non-zero only starting at cubic order. This is
\be
\psi(\vec x,t)=\eta(\vec x,t)-\frac{2}{7} s^2(\vec x,t)+\frac{4}{21}\delta(\vec x,t)^2\ .
\ee
This is all we need as in this paper we stop at one loop. For notational convenience, we can define non-vanishing products of the following variables as
\bea
&&s^2(\xfl,t)= s_{ij}(\xfl,t')  s^{ij}(\xfl,t')\ , \quad s^3(\xfl,t)=s_{ij}(\xfl,t')  s^{il}(\xfl,t')s_l{}^{j}(\xfl,t')\ ,\\ \nonumber
&& st(\xfl,t)=s_{ij}(\xfl,t')  t^{ij}(\xfl,t')\ , \quad \epsilon s(\xfl,t)=\epsilon_{ij}(\xfl,t')s^{ij}(\xfl,t'), \quad \epsilon t(\xfl,t)=\epsilon_{ij}(\xfl,t')t^{ij}(\xfl,t')\ ,
\eea
where indexes are lowered and raised with $\delta^{ij}$ and $\delta_{ij}$. By absorbing the trace into $\epsilon$, we can consider~$\epsilon_{ij}$ as traceless.
With these new variables, we can redefine the bias coefficients, up to cubic order, as
\bea\label{eq:euler_bias_3}
&&\delta_M(\vec x,t)\simeq \int^t dt'\; H(t')\; \left[   \bar c_{\delta}(t,t')\; :\delta(\xfl,t'): \right.\\  \nonumber
&&\qquad+\bar c_{\delta^2}(t,t')\;  :\delta(\xfl,t')^2:  +\bar c_{s^2}(t,t')\;  :s^2(\xfl,t'):\\\nonumber
&&\qquad+\bar c_{\delta^3}(t,t')\;   :\delta(\xfl,t')^3 : +\bar c_{\delta s^2}(t,t')\;   : \delta(\xfl,t') s^2(\xfl,t'):+\bar c_{\psi}(t,t')\;   :\psi(\xfl,t'):\\ \nonumber
&&\qquad\qquad+\bar c_{\delta st}(t,t') \;   :\delta(\xfl,t') st(\xfl,t'):+\bar c_{ s^3}(t,t')\;     :s^3(\xfl,t'):\\\nonumber
&&\qquad+ \bar c_{\epsilon}(t,t')\;\epsilon(\xfl,t')\\ \nonumber
&&\qquad+\bar c_{\epsilon\delta}(t,t') \;:\epsilon(\xfl,t')\delta(\xfl,t'):+\bar c_{\epsilon  s}(t,t') \;:\epsilon s(\xfl,t'):+\bar c_{\epsilon  t }(t,t') \;:\epsilon t(\xfl,t'):\\\ \nonumber
&&\qquad+\bar c_{\epsilon^2\delta}(t,t') \;:\epsilon(\xfl,t')^2\delta(\xfl,t'):+\bar c_{\epsilon\delta^2}(t,t') \;:\epsilon(\xfl,t')\delta(\xfl,t')^2:+\bar c_{\epsilon s^2}(t,t') \;:\epsilon(\xfl,t')s^2(\xfl,t'): \\ \nonumber
&&\qquad\qquad+\bar c_{\epsilon s \delta}(t,t') \;:\epsilon s(\xfl,t')\delta(\xfl,t'):+\bar c_{\epsilon  t \delta}(t,t') \;:\epsilon t(\xfl,t')\delta(\xfl,t'):\\ \nonumber
&&\left.\qquad+  \bar c_{\d^2\delta}(t,t')\;    \;\frac{\d^2_{x_{\rm fl}}}{\km^2}\delta(\xfl,t')+\dots\ \right] ,
\eea
In the first line we have the terms linear in the linear long wavelength fields,  in the second the quadratic ones, in the third and fourth the cubic ones. Starting from the fifth line, we have terms involving the stochastic bias. At the next line we can see that we are allowed to include a stochastic bias which  carries tensor indexes. Finally, in the last line, we have inserted the leading higher derivative term. The double columns before and after an operator, as for example $:\delta^2:$, means that we remove the expectation value of that operator itself. For example $:\delta(\vec x,t)^2:=\delta(\vec x,t)^2-\langle\delta(\vec x,t)^2\rangle$.

If we expand as usual the fluctuations as $\delta(\vec x,t)=\sum_n \delta(\vec x,t)^{(n)}$, and we assume that the time dependence of each fluctuation goes as it would in the EdS universe case, with the replacement $a\to D$, $\delta^{(n)}\sim \theta^{(n)}\sim D^n$, which is an approximation that seems to be valid at better than percent value in our universe~\cite{Carrasco:2012cv}, we can perform the time integrals in (\ref{eq:euler_bias_3}), and write, in Fourier space,
\bea\label{eq:euler_bias_4}
&&\delta_M(k,t)\quad=\quad  \\ \nonumber
&&\quad  =c_{\delta,1}(t) \; \delta^{(1)}(k,t)+ c_{\delta,2}(t)\; \delta^{(2)}(k,t)+ c_{\delta,3}(t)\delta^{(3)}(k,t)+c_{\delta,3_{c_s}}(t)\delta^{(3)}_{c_s}(k,t)\\ \nonumber
&&\quad+ \left[c_{\delta,1}(t)-c_{\delta,2}(t)\right]\; [\d_i \delta^{(1)}\; \frac{\d^i}{\d^2}\theta^{(1)}]_k(t)+ \left[c_{\delta,2}(t)-c_{\delta,3}(t)\right]\; [\d_i \delta^{(2)}\;\frac{\d^i}{\d^2}\theta^{(1)}]_k(t)\\ \nonumber
&&\qquad+ \frac{1}{2}\left[c_{\delta,1}(t)-c_{\delta,3}(t)\right]\; [\d_i \delta^{(1)}\;\frac{\d^i}{\d^2}\theta^{(2)}]_k(t)  \\ \nonumber
&&\quad+\left[ \frac{1}{2} c_{\delta,1}(t)- c_{\delta,2}(t)+\frac{1}{2} c_{\delta,3}(t)\right]\;\\ \nonumber
&&\qquad \times\left[ [\d_i \delta^{(1)}\; \frac{\d_j\d^i}{\d^2}\theta^{(1)}\; \frac{\d^j}{\d^2}\theta^{(1)}]_k(t)+[\d_i\d_j \delta^{(1)}\;\frac{\d^i}{\d^2}\theta^{(1)}\frac{\d^j}{\d^2}\theta^{(1)}]_k(t)\right]+ \\ \nonumber
&&\qquad+c_{\delta^2,1}(t) \; [\delta^2]^{(2)}_k(t)+c_{\delta^2,2}(t) \; [\delta^2]^{(3)}_k(t)-2 \left[c_{\delta^2,1}(t)-c_{\delta^2,2}(t)\right][\delta^{(1)}\d_i \delta^{(1)}\frac{\d^i}{\d^2}\theta^{(1)}]_k \\ \nonumber
&& \qquad
+c_{s^2,1}(t) \; [s^2]_k^{(2)}(t)+c_{s^2,2}(t) \; [s^2]_k^{(3)}(t)-2 \left[c_{s^2,1}(t)-c_{s^2,2}(t)\right][s_{lm}^{(1)}\d_i (s^{lm})^{(1)}\frac{\d^i}{\d^2}\theta^{(1)}]_k\\ \nonumber
&&\qquad +c_{st,1}(t) \; [st]^{(3)}_k(t)+ c_{\psi,1}(t)\;  \psi^{(3)}(k,t) + c_{\delta^3}[\delta^3]^{(3)}_k(t)+c_{\delta\,s^2}[\delta s^2]^{(3)}_k(t)\\ \nonumber
&&\qquad\quad+c_{s^3}[s^3]^{(3)}_k(t)+c_{\delta\,\epsilon^2}[\delta \epsilon^2]^{(3)}_k(t)\\ \nonumber
&&\qquad+ c_{\epsilon,1}(t) \;[\epsilon^{(1)}]_k+ c_{\epsilon,2}(t) \;[\epsilon^{(2)}]_k+\ldots\ .
\eea
In~(\ref{eq:euler_bias_4}), the superscript ${}^{(n)}$ refers to the order in initial fluctuations at which a certain quantity needs to be computed.  $\delta^{(3)}_{c_s}$ stays for the term obtained in the dar matter fluctuations after inserting the speed of sound counterterm at linear level. 
In particular, the various bias coefficient are given by
\be
c_{\delta,1}(t)= \int^t dt'\; H(t')\;     \bar c_{\delta}(t,t') \frac{D(t')}{D(t)}\ , \qquad c_{\delta,2}(t)= \int^t dt'\; H(t')\;     \bar c_{\delta}(t,t') \frac{D(t')^2}{D(t)^2}\ ,\quad  \ldots\ ,
\ee
that is by integrating the time-dependent bias-kernels with the time-dependence of the term we are considering. We see that the non-local in time bias can be thought as of having a different local bias for each different term in the perturbative expansion. We have distinguished $\epsilon^{1,2,\ldots}$ to distinguish the various contributions of the stochastic terms in a derivative expansions ($k^0,\,k^2,\,\ldots$), as they  will have in general different time dependence and so will lead to different local biases.

The terms in the lines from two to five in~(\ref{eq:euler_bias_4}) derive from the  Taylor expansion of the $\xfl$ variable inside the argument of the long wavelength $\delta(\xfl,t)$ appearing in the linear bias:
\bea\label{eq:xfl_expansion}
&&\delta(\xfl(\tau,\tau'),\tau')=\delta(\vec x,\tau')-\d_i\delta(x,\tau')\int_{\tau'}^\tau d\tau'' \; v^i(\vec x,\tau'')\\ \nonumber
&&\quad \quad\quad\quad\quad\quad+\frac{1}{2}\d_i\d_j \delta(x,\tau')\int_{\tau'}^\tau d\tau''\; v^i(\vec x,\tau'')\int_{\tau'}^\tau d\tau'''\;v^j(\vec x,\tau''')\\ \nonumber
&&\quad \quad\quad \quad\quad \quad+\d_i\delta(x,\tau')\int_{\tau'}^\tau d\tau''\; \d_j v^i(\vec x,\tau'') \int^\tau_{\tau''} \tau'''\;v^j(\vec x,\tau''')+\ldots\ .
\eea
We derive the expression for those terms in App.~\ref{app:fluid_bias}. Notice that their bias coefficients are related to the ones of $\delta^{(2)}$ and $\delta^{(3)}$, as this will be important for the cancellation of the IR divergencies in equal time correlations. Similar is the origin of the last terms in lines six and seven.

Let us now analyze the cross correlation between galaxies and dark matter. We will list all the terms that are larger than the two-loop contribution. By using~(\ref{eq:dark_matter_galaxy_expansion}), we have the following.

\begin{enumerate}

\item\label{point1} The first term, that we label by $_a$, comes from using the linear bias. It gives 
\bea\label{eq:point1}
&&\langle\delta_M(k,t)\delta(k,t)\rangle'_a\quad=\quad  \int^t dt'\; H(t')\; \left[   \bar c_{\delta}(t,t')\; \langle[\delta(\xfl,t')]_k\delta(k,t)\rangle'_\text{EFT-one-loop} \right]\\ \nonumber
&&\quad  =c_{\delta,1}(t) \; \langle\delta^{(1)}(k,t)\delta^{(1)}(k,t)\rangle'+ c_{\delta,2}(t)\; \langle\delta^{(2)}(k,t)\delta^{(2)}(k,t)\rangle'\\ \nonumber
&&\quad+\left[ c_{\delta,3}(t)+ c_{\delta,1}(t)\right]\; \langle\delta^{(3)}(k,t)\delta^{(1)}(k,t)\rangle'+\left[c_{\delta,3_{c_s}}(t)+c_{\delta,1}(t)\right]\; \langle\delta^{(3)}_{c_s}(k,t)\delta^{(1)}(k,t)\rangle'\ \\ \nonumber
&&\quad+ \left[c_{\delta,1}(t)-c_{\delta,2}(t)\right]\; \langle[\d_i \delta^{(1)}\; \frac{\d^i}{\d^2}\theta^{(1)}]_k(t)\delta^{(2)}(k,t)\rangle' \\ \nonumber
&&\qquad+ \left[c_{\delta,2}(t)-c_{\delta,3}(t)\right]\; \langle[\d_i \delta^{(2)}\;\frac{\d^i}{\d^2}\theta^{(1)}]_k(t)\delta^{(1)}(k,t)\rangle'\\ \nonumber
&&\qquad+ \frac{1}{2}\left[c_{\delta,1}(t)-c_{\delta,3}(t)\right]\; \langle[\d_i \delta^{(1)}\;\frac{\d^i}{\d^2}\theta^{(2)}]_k(t)\delta^{(1)}(k,t)\rangle'  \\ \nonumber
&&\quad+\left[ \frac{1}{2} c_{\delta,1}(t)- c_{\delta,2}(t)+\frac{1}{2} c_{\delta,3}(t)\right]\;\\ \nonumber
&&\qquad \times\left[ \langle[\d_i \delta^{(1)}\; \frac{\d_j\d^i}{\d^2}\theta^{(1)}\; \frac{\d^j}{\d^2}\theta^{(1)}]_k(t)\delta^{(1)}(k,t)\rangle'+\langle[\d_i\d_j \delta^{(1)}\;\frac{\d^i}{\d^2}\theta^{(1)}\frac{\d^j}{\d^2}\theta^{(1)}]_k(t)\delta^{(1)}(k,t)\rangle'\right] ,
\eea
where we have used that 
\be
\langle\delta^{(3)}(k,t)\delta^{(1)}(k,t)\rangle'= \langle\delta^{(1)}(k,t)\delta^{(3)}(k,t)\rangle'\ , \; \langle\delta^{(3)}_{c_s}(k,t)\delta^{(1)}(k,t)\rangle'= \langle\delta^{(1)}(k,t)\delta^{(3)}_{c_s}(k,t)\rangle'\ ,\; \ldots\ .
\ee
Notice that the bias coefficients of the terms arising from the expansion of $\xfl$ are related to the ones in the second line for the $\langle\delta^{(3)}\delta^{(1)}\rangle'$ and $\langle\delta^{(2)}\delta^{(2)}\rangle'$ contributions. This could have been expected from the fact that IR divergencies need to cancel, and neither the $\langle\delta^{(3)}\delta^{(1)}\rangle'$ nor the $\langle\delta^{(2)}\delta^{(2)}\rangle'$ diagrams are IR safe~\footnote{It is straightforward to see that thanks to the fact that the dark matter field is evaluated on $\xfl$, we have that our expression for the equal time dark matter galaxies correlation is IR safe, as it should be for symmetry reasons. Under a time-dependent, spatially constant diffeomorphism, we have
\be
x^i\quad\to\quad x^i+\int_0^\tau d\tau'\; V^i(\tau')\ , \qquad v^i(\tau)\quad\to\quad v^i(\tau)+V^i(\tau)\ .
\ee
This leads to the following transformations for the dark matter field and the galaxy field (for simplicity, we perform the transformation for the galaxy field only at leading order in the spatial dependence of the velocity).  Notice first that in Fourier space we have
\be
\delta_M(\vec k_1,\tau_1)=\int_0^{\tau_1} d\tau_1'\; a(\tau_1')\; C(\tau_1,\tau_1')\; \delta(\vec k_1,\tau_1)\; e^{-i\vec k_1^i\cdot\int^{\tau_1}_{\tau_1'}d\tau_1'' v^i(\vec k_1,\tau''_1)}\ .
\ee
We therefore have
\bea
&&\delta(\vec k_1,\tau_1)\quad\to\quad \delta(\vec k_1,\tau_1) \; e^{- i\, k_1^i\cdot\int_0^{\tau_1} d \tau_1'\; V^i(\tau_1') } \ ,\\ \nonumber
&&\delta_M(\vec k_1,\tau_1)\quad\to\quad \int^{\tau_1}_0 d\tau_1'\;a(\tau_1')\; C(\tau_1,\tau_1')\;\delta(\vec k_1,\tau_1) e^{-i\,\vec k_1^i\cdot\int^{\tau_1}_{0}d\tau_1'' v^i(\vec k_1,\tau''_1)} \; e^{- i\, k_1^i\cdot\int_0^{\tau_1} d \tau_1''\; V^i(\tau_1'') } \\ \nonumber
&&\qquad\qquad\qquad\qquad =\delta_M(\vec k_1,\tau_1) \;  e^{- i\, k_1^i\cdot\int_0^{\tau_1} d \tau_1'\; V^i(\tau_1') }\ .
\eea
This tells us that $\langle\delta_M(k,\tau)\delta(k,\tau)\rangle'$ and $\langle\delta_M(k,\tau)\delta_M(k,\tau)\rangle'$ are invariant under such a diff., which means that they are IR safe.
}.
We see that the fact that the linear bias is non-local in time is equivalent to include three different local biases. 

\item In a one-loop computation, the second term can just be evaluated at tree level. It therefore gives
\bea
&&\langle\delta_M(k,t)\delta(k,t)\rangle'_b\quad=\quad  \int^t dt'\; H(t')\; \left[   \bar c_{\delta^2}(t,t')\; \langle[\delta^2]_{k}(t')\delta(k,t)\rangle'_\text{EFT-one-loop} \right]=\\ \nonumber
&&\quad  =c_{\delta^2,1}(t) \; \langle[\delta^2]^{(2)}_k(t)\delta^{(2)}(k,t)\rangle'+c_{\delta^2,2}(t) \; \langle[\delta^2]^{(3)}_k(t)\delta^{(1)}(k,t)\rangle'\\ \nonumber
&&\qquad-2 \left[c_{\delta^2,1}(t)-c_{\delta^2,2}(t)\right]\langle[\delta^{(1)}\d_i \delta^{(1)}\frac{\d^i}{\d^2}\theta^{(1)}]_k(t)\delta^{(1)}(k,t)\rangle'\ .
\eea

\item The third term is very similar to the second one. It gives
\bea
&&\langle\delta_M(k,t)\delta(k,t)\rangle'_c\quad=\quad  \int^t dt'\; H(t')\; \left[   \bar c_{s^2}(t,t')\; \langle[s^2]_{k}(t')\delta(k,t)\rangle'_\text{EFT-one-loop} \right]\\ \nonumber
&&\quad  =c_{s^2,1}(t) \; \langle[s^2]_k^{(2)}(t)\delta^{(2)}(k,t)\rangle'+c_{s^2,2}(t) \; \langle[s^2]_k^{(3)}(t)\delta^{(1)}(k,t)\rangle'\\ \nonumber 
&&\qquad-2 \left[c_{s^2,1}(t)-c_{s^2,2}(t)\right]\langle[s_{lm}^{(1)}\d_i (s^{lm})^{(1)}\frac{\d^i}{\d^2}\theta^{(1)}]_k(t)\delta^{(1)}(k,t)\rangle'\ . 
\eea

\item Let us now pass to the cubic biases. 
We have 
\bea\nonumber
&&\langle\delta_M(k,t)\delta(k,t)\rangle'_d\quad=\quad \int^t dt'\; H(t')\; \\ \nonumber
&&  \qquad\times \left[   \bar c_{st}(t,t')\; \langle[st]_{k}(t')\delta(k,t)\rangle'+\bar c_{\psi}(t,t')\;\langle\psi(k,t')\delta(k,t)\rangle'\right. \\ \nonumber
&&\qquad  \quad\bar c_{\delta^3}(t,t')\; \langle[\delta^3]_{k}(t')\delta(k,t)\rangle'+\bar c_{\delta\,s^2}(t,t')\; \langle[\delta\,s^2]_{k}(t')\delta(k,t)\rangle'+\\ \nonumber
&&\qquad\quad\left.\bar c_{\delta \epsilon^2}(t,t')\; \langle[\delta\, \epsilon^2]_{k}(t')\delta(k,t)\rangle'+\bar c_{s^3}(t,t')\; \langle[s^3]_{k}(t')\delta(k,t)\rangle' \right]_\text{EFT-one-loop} \\ 
&& \qquad =c_{st,1}(t) \; \langle[st]^{(3)}_k(t)\delta^{(1)}(k,t)\rangle'+ c_{\psi,1}(t)\; \langle \psi^{(3)}(k,t)\delta^{(1)}(k,t)\rangle'\\ \nonumber
&&\qquad\quad + c_{\delta^3}\langle[\delta^3]^{(3)}_k(t)\delta^{(1)}(k,t)\rangle'+c_{\delta\,s^2}\langle[\delta s^2]^{(3)}_k(t)\delta^{(1)}(k,t)\rangle'\\ \nonumber
&&\qquad\quad+c_{s^3}\langle[s^3]^{(3)}_k(t)\delta^{(1)}(k,t)\rangle'+c_{\delta\,\epsilon^2}\langle[\delta \epsilon^2]^{(3)}_k(t)\delta^{(1)}(k,t)\rangle'\ .
\eea

\item The leading spatial derivative term is also important at one-loop level.
\bea\nonumber
&&\langle\delta_M(k,t)\delta(k,t)\rangle'_e\quad=\quad \int^t dt'\; H(t')\;\left[   \bar c_{\d^2\delta}(t,t')    \;\frac{\d^2_{x_{\rm fl}}}{\km^2}\langle\delta(\xfl,t')\delta(\vec x,t)\rangle'_\text{EFT-one-loop}\right]_k  \\ 
&& \qquad =c_{\d^2\delta,1}(t) \;\frac{k^2}{k_{\rm M}^2} \langle\delta^{(1)}(k,t)\delta^{(1)}(k,t)\rangle'\ .
\eea

\item The leading stochastic contribution scales as
\bea\nonumber
&&\langle\delta_M(k,t)\delta(k,t)\rangle'_f\quad=\quad \int^t dt'\; H(t')\;   \bar c_{\epsilon}(t,t')    \;\langle\epsilon(k,t')[\d^2\Delta\tau]_k(t)\rangle'_\text{EFT-one-loop}  \\  \nonumber
&& \qquad =c_{\epsilon,1}(t) \;\frac{\gamma^{1/2}}{(\km^3\knl^3)^{1/2}}\cdot\frac{k^2}{\knl^2}\ ,
\eea
where we took 
\be
\langle[\epsilon]_k[\d^2\Delta\tau]_k\rangle'= \frac{\gamma^{1/2}}{(\km^3\knl^3)^{1/2}}\cdot\frac{k^2}{\knl^2}\ .
\ee
This is too an high power in $k$ to make it important at one-loop order (not quite so at higher orders), so we will neglect it here.

\end{enumerate}

In summary, in the non-local in time case, for the dark matter galaxies cross correlation there are several parameters to be measured in simulations or in observations.  The large number of parameters is a reason of concern for what the predictive capability of the theory is. Though it looks like that there are indeed many parameters that are necessary, there are several ways in which the actual number of parameters necessary to fit an observation is reduced: some of these coefficients might give rise to a very similar functional forms; some might appear in several different observable, such as, as we will see, in the power spectrum and the bispectrum; finally, the time-local approximation might reduce the number of needed parameters. We will explore this next.

 \subsection{Time-local Approximation}
 
 For the case of the dark matter power spectrum at two loops~\cite{Carrasco:2013mua}, it was shown that the local in time treatment was a good approximation to the numerical data. As we described in~Sec.~\ref{sec:quasi-local}, this is now understood in terms of the fact that the theory can be treated as local to some degree of approximation, and that in the two-loop dark matter power spectrum the difference between a local and a non-local treatment started mattering only at the highest term computed in perturbation theory.

In the case of collapsed objects there is a difference, as the difference between local in time and non-local in time appears earlier in perturbation theory.  In particular, the first terms that are affected by this difference are the one-loop terms, contrary to the two-loop terms as in the case of the dark matter power spectrum. Similarly to what we just discussed, this tells us that for a one-loop calculation one can approximate the calculation for the biased objects as local in time: differences will be about 1/10 of the one-loop result, which is about the two-loop term.

This simplifies remarkably the one-loop formulas of the former sec.~\ref{sec:dark-matter-cross}. In the local in time approximation we obtain the following.

Terms $a,b,c$ change as
\bea\label{eq:point1local}
&&\langle\delta_M(k,t)\delta(k,t)\rangle'_a=\quad  \int^t dt'\; H(t')\; \left[   \bar c_{\delta}(t,t')\; \langle[\delta(\xfl,t')]_k\delta(k,t)\rangle'_\text{EFT-one-loop} \right]=\\ \nonumber
&&\qquad  \simeq c_{\delta,\text{local},1}(t) \;\left[ \langle\delta(k,t)\delta(k,t)\rangle'_\text{tree}+ \langle\delta(k,t)\delta(k,t)\rangle'_\text{1-loop}+ \langle\delta(k,t)\delta(k,t)\rangle'_{c_s,k^2 P_{11}}\right]\ , \\ \nonumber
&&\langle\delta_M(k,t)\delta(k,t)\rangle'_b\quad=\quad  \int^t dt'\; H(t')\; \left[   \bar c_{\delta^2}(t,t')\; \langle[\delta^2]_{k}(t')\delta(k,t)\rangle'_\text{EFT-one-loop} \right]=\\ \nonumber
&&\qquad  \simeq c_{\delta^2,\text{local},1}(t) \; \left[\langle[\delta^2]^{(2)}_k(t)\delta^{(2)}(k,t)\rangle'+ \langle[\delta^2]^{(3)}_k(t)\delta^{(1)}(k,t)\rangle'\right]\ , \\ \nonumber
&&\langle\delta_M(k,t)\delta(k,t)\rangle'_c=\quad  \int^t dt'\; H(t')\; \left[   \bar c_{s^2}(t,t')\; \langle[s^2]_{k}(t')\delta(k,t)\rangle'_\text{EFT-one-loop} \right]\\ \nonumber
&&\qquad  \simeq c_{s^2,\text{local},1}(t) \left[\langle[s^2]_k^{(2)}(t)\delta^{(2)}(k,t)\rangle'+ \langle[s^2]_k^{(3)}(t)\delta^{(1)}(k,t)\rangle'\right]\ ,
\eea
while the $d,e,f$ terms remain unchanged, apart for some obvious rescaling by $1/\oms$.
Notice that because of the quasi time locality, we should add a term of the form
\be\label{eq:time_derivative}
c_{D_t\delta,\text{local},1}(t) \frac{1}{\oms}\langle[\d_t\delta+v^i\d_i\delta]_k(t)\delta(k,t)\rangle'
\ee
This term comes from the flow time derivative, and originates from the quasi-local in time expansion of the $\xfl$ dependence. However, this term can be neglected at one-loop order. The term in $\d_t\delta^{(1)}(k,t)\delta^{(1)}(k,t)$ is degenerate with the linear bias term because $\d_t\delta^{(1)}\propto \delta^{(1)}$. Explicitly, one can redefine the linear bias as $ c_{\delta,\text{local},1}(t)\to  c_{\delta,\text{local},1}(t)+c_{D_t\delta,\text{local},1}(t)\frac{\H}{\oms} \frac{D'(a)}{D(a)}$, which is a small correction, so that term in $\d_t\delta$ appears in the linear combination $\d_t\delta-\frac{D'}{D} \delta$, that starts at second order. This implies that the contribution of this term is comparable to a one loop term times $\H/\oms$, which is about equivalent to a two-loop term, and so negligible. This kind of redefinitions are the same ones that allowed us to consider $t_{ij}$ and $\psi$ as independent variables in the bias expansion. The term $\langle[v^i\d_i\delta]_k(t)\delta(k,t)\rangle'$ is a one loop term suppressed by $\H/\oms$, and so it is about the size of a two loop term as well. The combination in~(\ref{eq:time_derivative}) can be easily derived assuming the the kernel $c_\delta(t,t')$ is peaked around $t$ with width of order $1/\oms$, and it originates from the time dependence and the $\xfl$ dependence of the variable $\delta(\xfl, t)$. We show this explicitly in App.~\ref{eq:flow-time-derivative}.  Notice that now there is only one bias coefficients originating from the linear and the quadratic bias.

We see that in the local in time approximation, there are much less bias coefficients that are needed. Notice however that some of the terms we have neglected at one loop will become important in a calculation done at two loops, even in the local in time approximation, where the dark matter power spectrum has been shown to agree with simulations up to $k\simeq 0.6 \hinvMpc$. In a sense, the calculation will become much more interesting, but also more complex, at this order. We leave this to future work.

\subsection{Explicit Expressions and Renormalization}

For completeness, we give the precise expressions of the terms we described before. This will allow us to explicitly verify the IR-safety of the equal time cross correlation and to discuss renormalization. The calculation can be carried out using the standard techniques in SPT. We have the following.

\begin{enumerate}

\item The $a$ contribution gives 
\bea\label{eq:pointa}
&&\langle\delta_M(k,t)\delta(k,t)\rangle'_a\quad=\quad \\ \nonumber
&&\ \  =c_{\delta,1}(t) P_{11}(k;t,t)+c_{\delta,2}(t) P_{22}(k;t,t)+\left[ c_{\delta,3}(t)+ c_{\delta,1}(t)\right] P_{31}(k;t,t)\\ \nonumber
&&\quad+\left[c_{\delta,3_{c_s}}(t)+c_{\delta,1}(t)\right]P_{11,c_s}(k;t,t)+\\ \nonumber
&&\quad+ \left[c_{\delta,1}(t)-c_{\delta,2}(t)\right] \int \frac{d^3q}{(2\pi)^3}\; 2 \;\frac{(\vec k-\vec q)\cdot \vec q}{q^2} F_{2,S}(-\vec q,\vec q-\vec k) \, P_{11}(q;t,t)\,P_{11}(|\vec k-\vec q|;t,t) \\ \nonumber
&&\quad+ \left[c_{\delta,2}(t)-c_{\delta,3}(t)\right] \int \frac{d^3q}{(2\pi)^3} \;2\; \frac{(\vec k-\vec q)\cdot \vec q}{q^2} F_{2,S}(-\vec q,\vec k) \, P_{11}(q;t,t)\,P_{11}(k;t,t) \\ \nonumber
&&\quad+ \left[c_{\delta,1}(t)-c_{\delta,3}(t)\right] \int \frac{d^3q}{(2\pi)^3}\;2\, \frac{(\vec k-\vec q)\cdot \vec q}{q^2}  G_{2,S}(\vec k,-\vec q) \, P_{11}(q;t,t)P_{11}(k;t,t)  \\ \nonumber
&&\quad -\left[ \frac{1}{2} c_{\delta,1}(t)- c_{\delta,2}(t)+\frac{1}{2} c_{\delta,3}(t)\right] 
\,\frac{1}{3}\,\int \frac{d^3q}{(2\pi)^3}  \left(1+\frac{k^2}{q^2}\right) \, P_{11}(q;t,t)\,P_{11}(k;t,t)\  ,
\eea
where 
\be
P_{ij}(k;t_1,t_2)=\langle\delta^{(i)}(\vec k,t_1)\delta^{(j)}(\vec k,t_2)\rangle'\ ,
\ee
and where we have performed the angular integrations that could be done analytically~\footnote{The definitions of $F_{n,S}$ and $G_{n,S}$ are the standard ones of SPT. We expand $\delta$ and $\theta$ as
\bea
\label{eq:delansatz}
&&\delta(a,\vk) = \sum_{n=1}^\infty \, [D_1(a)]^n \delta^{(n)}(\vk)\ , \\
&&\d_iv^i(a,\vk)=-\theta(a,\vk) =- \H(a) f \sum_{n=1}^\infty \, [D_1(a)]^n \theta^{(n)}(\vk))\ ,
\eea
and then we write the $\vk$-dependent solutions in the following manner:
\bea
\delta^{(n)}(\vk) &=& \int \! \frac{d^3 \vq_1}{(2\pi)^3} \cdots \int \! \frac{d^3 \vq_n}{(2\pi)^3}
(2\pi)^3 \delta_{\rm D}(\vk-\vq_{1\cdots n})
F_n(\vq_1,\dots,\vq_n)
\delta(\vq_1)\cdots\delta(\vq_n)\ , \\
\theta^{(n)}(\vk) &=& \int \! \frac{d^3 \vq_1}{(2\pi)^3} \cdots \int \! \frac{d^3 \vq_n}{(2\pi)^3}
(2\pi)^3 \delta_{\rm D}(\vk-\vq_{1\cdots n})
G_n(\vq_1,\dots,\vq_n)
\delta(\vq_1)\cdots\delta(\vq_n)\ ,
\label{eq:thetakansatz}
\eea
$F_{n,S}$ and $G_{n,S}$  are the symmetryzed versions of $F_{n}$ and $G_{n}$. 
}. 
By using this expression, we have checked that the IR divergencies do cancel out.

\item The $b$ contribution gives 
\bea
&&\langle\delta_M(k,t)\delta(k,t)\rangle'_b\quad=\\ \nonumber
&&\quad  =c_{\delta^2,1}(t) \; \int \frac{d^3q}{(2\pi)^3}\;  2\, F_{2,S}(\vec k-\vec q,\vec q)\;P_{11}(q;t,t)\,P_{11}(|\vec k-\vec q|;t,t) \\ \nonumber
&&\qquad +c_{\delta^2,2}(t) \; 2\times \frac{34}{21} \sigma^2(t) P_{11}(k;t,t)\\ \nonumber
&&\qquad +2\left[c_{\delta^2,1}(t)-c_{\delta^2,2}(t)\right]  \sigma^2(t) P_{11}(k;t,t)\ ,
\eea
where
\be
\sigma^2(t)=\langle\delta(\vec x,t)^2\rangle=\int \frac{d^3q}{(2\pi)^3}\; P_{11}(q;t,t)\ .
\ee

\item The $c$ contribution gives  
\bea
&&\langle\delta_M(k,t)\delta(k,t)\rangle'_c\quad=\\ \nonumber
&&\quad  =c_{s^2,1}(t) \; \int \frac{d^3q}{(2\pi)^3}\;  2\, F_{2,S}(\vec k-\vec q,\vec q)\; S(\vec k-\vec q,\vec q)\;P_{11}(q;t,t)\,P_{11}(|\vec k-\vec q|;t,t) \\ \nonumber
&&\qquad +c_{s^2,2}(t) \;\left[ \int \frac{d^3q}{(2\pi)^3}\; 4\, F_{2,S}(\vec k,-\vec q)\; S(\vec k-\vec q,\vec q)\;P_{11}(q;t,t)\right]P_{11}(k;t,t)\\ \nonumber
\eea
where 
\be
S(\vec k,\vec q)=\frac{(\vec q\cdot\vec k)^2}{k^2 q^2}-\frac{1}{3}\ .
\ee
and where we have found that the coefficient of $\left[c_{s^2,1}(t)-c_{s^2,2}(t)\right]$ vanishes after angular integration.

\item The $d$ contribution gives 
\bea
&&\langle\delta_M(k,t)\delta(k,t)\rangle'_d\quad=\\ \nonumber
&&\quad  =c_{s t,1}(t) \; \int \frac{d^3q}{(2\pi)^3}\;  2\; S(\vec k-\vec q,\vec q)\; D_{2,S}(-\vec q,\vec k)\;P_{11}(q;t,t)\,P_{11}(\vec k;t,t) \\ \nonumber
&&\qquad +c_{\psi,1}(t) \int \frac{d^3q}{(2\pi)^3}\;  \left[3\, D_{3,S}(\vec k,\vec q,-\vec q)-4\,D_{2,S}(\vec q,\vec k-\vec q) F_{2,S}(-\vec q,\vec k) \right]\;P_{11}(q;t,t)\,P_{11}(\vec k;t,t)\\ \nonumber
&&\qquad +\left[c_{\delta^3,1}(t) 3 \sigma^2(t)+c_{\delta s^2,1}(t) \frac{2}{3} \sigma^2(t)+\sigma_\epsilon^2(t)\right]P_{11}(\vec k;t,t) \ ,
\eea
where
\be
D_{n,S}(\ldots)=G_{n,S}(\ldots)-F_{n,S}(\ldots)\  , 
\ee
and
\be
\sigma_\epsilon^2(t)=\langle\epsilon(\vec x,t)^2\rangle\ .
\ee

\item Then the $e$ contribution gives 
\be
\langle\delta_M(k,t)\delta(k,t)\rangle'_e\quad=\quad c_{\d^2 \delta,1} \frac{k^2}{\km^2} P_{11}(\vec k;t,t)\ .
\ee

\item Finally, the $f$ contribution gives
\bea\nonumber
&&\langle\delta_M(k,t)\delta(k,t)\rangle'_f\quad =c_{\epsilon,1}(t) \;\frac{\gamma^{1/2}}{(\km^3\knl^3)^{1/2}}\cdot\frac{k^2}{\knl^2}\ ,
\eea

\end{enumerate}

It can be checked that IR divergencies cancel explicitly, which is a consistency check of our formulas. 

\subsubsection{UV dependence and renormalization}

When doing calculation in any field theory, be it either classical or quantum, it is important to make our predictions insensitive to the mistake done by using some equations in a regime in which they are not valid. In the case of LSS, we do not have control on the physics at distances shorter than the non-linear scale. The whole point of the EFTofLSS is to make our predictions insensitive to what we assume for the physics at such short distances. Therefore, when describing collapsed objects, we have to insure that our predictions are insensitive to the same regime.
 The calculation of the dark matter power spectrum at two loops seem to show that dark matter can be predicted up to the scale $k\simeq 0.6\hinvMpc$. It is therefore fair to take the point of view that physics is perturbative up to that scale, while we want to be insensitive to what happens at scales shorter than that. The former one-loop expressions for the galaxies dark matter cross correlation involve convolution integrals that can potentially be sensitive to the behavior of the dark matter power spectrum on scales shorter than $k\simeq 0.6\hinvMpc$. The contribution from that regime is not under control, and we must therefore ensure that we are not sensitive to that. This can be achieved by readjusting the bias parameter to cancel the contribution from that regime and ensure that we can obtain the correct result.  This was described in~\cite{McDonald:2009dh}, and very recently reemphasized in great detail in~\cite{Assassi:2014fva}. At around $k\simeq 0.6\hinvMpc$, the matter power spectrum can be approximately described by a scaling one with slope $n\simeq-2.1$. With this slope, let us identify, among the various contributions, the ones that are UV sensitive. This can be done by Taylor expanding in $q\gg k$ the integrands in the former expressions, and identifying the ones that are UV divergent if the power spectrum has a slope $n\simeq-2.1$. Let us find these contributions. 
 
 Let us start from the terms in $a$. After Taylor expansion, we identify the following terms that are UV divergent. We ignore the loop terms that appear already in the calculation of the dark matter correlation functions, as they are renormalized by the counterterms present in the equations for the dark matter~\footnote{To this end, one can notice that in case $P_{13}$ or $P_{22}$ are UV divergent, the  counterterms from the stress tensor, such as $c_s$ for example, are able to reabsorb those divergencies. In this case, it must be that the coefficients such as $c_{s}$ split into a counterterm part and a finite part, each one with its own time dependence, and therefore each generating its own bias coefficients.}. We therefore concentrate on the other terms. After Taylor expansion in $q\gg k$, and using that $n\simeq -2.1$, the only divergent terms are the leading ones in the Taylor expansion and lead to the following expression:
 \bea\label{eq:UVa}
&&\langle\delta_M(k,t)\delta(k,t)\rangle'_{a,\,\rm UV}\quad= \\ \nonumber
&&-\left\{\frac{34}{21}\left[c_{\delta,2}(t)-c_{\delta,3}(t)\right]-\frac{5}{21}\left[c_{\delta,1}(t)-c_{\delta,3}(t)\right]+\frac{1}{3}\left[\frac{1}{2}c_{\delta,1}(t)-c_{\delta,2}(t)+\frac{1}{2}c_{\delta,3}(t)\right]\right\}\;\sigma^2(t)\;  P_{11}( k;t,t).
\eea
Similarly, from the $\delta^2$ bias, we have
 \bea\label{eq:UVb}
&&\langle\delta_M(k,t)\delta(k,t)\rangle'_{b,\,\rm UV}\quad= \quad\left[\frac{68}{21} c_{\delta^2,2}(t)+2\left[c_{\delta^2,1}(t)-c_{\delta^2,2}(t)\right]\right]\;\sigma^2(t)\;  P_{11}(\vec k;t,t)\ .
\eea
 The $s^2$ bias gives
  \bea\label{eq:UVc}
&&\langle\delta_M(k,t)\delta(k,t)\rangle'_{c,\,\rm UV}\quad= \quad\frac{136}{63} c_{s^2,2}(t)\;\sigma^2(t)\;  P_{11}( k;t,t)\ .
\eea
 From the $d$ contribution we finally have
   \bea\nonumber\label{eq:UVd}
&&\langle\delta_M(k,t)\delta(k,t)\rangle'_{d,\,\rm UV}\quad= \quad\left\{\left[\frac{16}{63} c_{st,1}(t)+ 3 c_{\delta^3,1}(t)+ \frac{2}{3} c_{\delta s^2,1}(t)\right]\;\sigma^2(t)+\sigma_{\epsilon}^2(t)\right\}\;  P_{11}( k;t,t) \ . \\
\eea
The $e$ and $f$ terms, being tree level, do not include any UV sensitive term. 

In order to be able to correctly parametrize the UV-sensitivity from our expressions, we need to able to readjust one of our bias parameters to change this contribution to make it correct. This is possible because indeed the form of all of these UV divergent terms is the one of the linear bias. We can therefore renormalize the linear bias coefficient to absorb it. In practice, we imagine that $c_{\delta,1}$ is a bare parameter, sum of a finite part and a counterterm:
\be
c_{\delta,1}(t)=c_{\delta,1,\rm finite}(t)+c_{\delta,1,\rm counter}(t)\ ,
\ee
where  the counterterm is chosen so that the final answer is independent of the short distance physics:
\bea
&&c_{\delta,1,\rm counter}(t)=\\ \nonumber
&&\quad -\;\sigma^2(t)\left\{\frac{34}{21}\left[c_{\delta,2}(t)+c_{\delta,3}(t)\right]-\frac{5}{21}\left[c_{\delta,1,\rm finite}(t)-c_{\delta,3}(t)\right]+\frac{1}{3}\left[\frac{1}{2}c_{\delta,1,\rm finite}(t)-c_{\delta,2}(t)+\frac{1}{2}c_{\delta,3}(t)\right]\right.\\ \nonumber
&&\quad\quad\quad\left.+\frac{68}{21} c_{\delta^2,2}(t) +2\left[c_{\delta^2,1,\rm finite}(t)-c_{\delta^2,2}(t)\right]+\frac{136}{63} c_{\delta^2,2}(t)+\frac{16}{63} c_{st,1}(t)+ 3 c_{\delta^3,1}(t)+ \frac{2}{3} c_{\delta s^2,1}(t)\right\}\\ \nonumber
&&\quad-\sigma_{\epsilon}^2(t)\ .
\eea

Notice that, in our one-loop calculation, one should shift {\it only} the $c_{\delta,1}$ that multiplies the tree level result $P_{11}$. One  should {\it not} shift at this order the terms $c_{\delta,1}$ that multiply one-loop terms~\footnote{This is similar to that happens for the $c_s$ counterterm in the dark matter power spectrum at two-loops~\cite{Carrasco:2013mua}, where at two-loops the one-loop counterterm is evaluated with the one-loop $c_s$, while the tree level counterterm is evaluated with the two-loop $c_s$.}.  In practice, this procedure at one loop order amounts to removing the UV limits of each of the terms in $a,b,c,d$. 

In the procedure we have just described, there is an ambiguity that we still have to resolve. It has to do on how to identify the divergent part of the contribution, as in practice we can always add a finite part to a divergent part, and the divergent part is still divergent. The procedure to fix this ambiguity is to impose the overall prediction of the theory to agree with the observation at some specific wavenumber of interest, $k_{\rm ren}$, that goes under the name of renormalization point. Since perturbation theory is organized in powers of $k/\knl$ or $k/\km$, this procedure should be done at the lowest possible $k$'s so that the terms that have not been included in the calculation are negligible. In practice, since cosmic variance grows in the infrared, $k_{\rm ren}$ cannot be chosen arbitrarily small, and one has to resort to performing higher order calculations so that the dependence of the parameters on $k_{\rm ren}$ becomes negligible. This procedure has been applied to simulated data first in~\cite{Carrasco:2012cv}. Alternatively, one can measure the bias parameters directly in simulations, in a procedure similar to the one that was used in~\cite{Carrasco:2012cv} to measure the $c_s$ parameter of dark matter from small distance correlations~\footnote{We refer to~\cite{Carrasco:2012cv} for details on the procedure, which involves a few subtle points.}. Once one of this two procedures has been performed, the process of renormalization is concluded, and at this stage the result is independent of the UV contributions. 

After doing this, as it was notice in~\cite{McDonald:2009dh} where a subset of the terms that are present here are also present, the finite contributions of $c_{s^2,2},c_{\psi,1}$ and $c_{st,1}$ are all proportional to each other. This means that, at this one-loop order, they are degenerate and can be treated as one parameter. This is not expected to hold at higher orders. 

For illustrative proposes, let us explain a different, but completely equivalent, way to look at the renormalization of the bias coefficients. We were driven to perform a renormalization because our expressions were sensitive to UV physics. This can be traced back to the fact that our expressions contained product of operators at the same location. One example is the operator that appear in the quadratic bias $\delta(\vec x,t)^2$. Since this is a product of different operators at the same location, it is extremely sensitive to short distance physics. We already partially dealt with this when we defined it as $:\delta(\vec x,t)^2:=\delta(\vec x,t)^2-\langle\delta(\vec x,t)^2\rangle$, so that  its vacuum expectation value vanished. But this fixes the ambiguities only at some level. In fact, in order to correctly define this operator we need to define all its correlation functions with all the relevant long wavelength fields. This can be done by adding suitable counterterms. In our case, at one loop order, for all our operators we need only to be concerned with their correlation with the density $\delta(\vec x,t)$. The first example of this kind of composite, or contact, operators that appeared in the context of the EFTofLSS has been the dark matter velocity field $\vec v$, which is defined as the ratio at the same location of the momentum field and the density field: $\vec v(\vec x,t)=\vec \pi(\vec x,t)/\rho(\vec x,t)$. As discussed in~\cite{Carrasco:2013mua}, in order to compute correlation functions at one loop for the velocity field, one needs to define its correlation functions with $\delta$, and, in order to do that, one has to add a suitable counterterm~\cite{Carrasco:2013mua}
\be\label{eq:velocity_def_ren}
v_{R}^i(\vec x,t)=v^i(\vec x,t)-\int \frac{d a'}{a' \H'} \,K_{\rm v}(a,a')\, \d^i\delta(a', \vec x_{\rm fl})+\ldots\ ,
\ee
where the subscript $R$ stays for renormalized  and where $K_{\rm v}(a,a')$ is a suitable Kernel to be fixed by requiring that the UV sensitiveness is cancelled and $v^i$ correlation functions agree with the observed ones at low $k$'s. $\ldots$ represent higher derivative or higher order terms. At one loop, the first term is enough.  In the case of bias, we can do the same construction. We can rewrite (\ref{eq:euler_bias_4}) directly in terms of renormalized fields
\bea\label{eq:euler_bias_5}
&&\delta_M(k,t)\quad=\quad  \\ \nonumber
&&\quad  =c_{\delta,1}(t) \; \delta^{(1)}(k,t)+ c_{\delta,2}(t)\; \delta^{(2)}(k,t)+ c_{\delta,3}(t)\delta^{(3)}(k,t)+c_{\delta,3_{c_s}}(t)\delta^{(3)}_{c_s}(k,t)\\ \nonumber
&&\quad+ \left[c_{\delta,1}(t)-c_{\delta,2}(t)\right]\; [\d_i \delta^{(1)}\; \frac{\d^i}{\d^2}\theta^{(1)}]_{R,\, k}(t)+ \left[c_{\delta,2}(t)-c_{\delta,3}(t)\right]\; [\d_i \delta^{(2)}\;\frac{\d^i}{\d^2}\theta^{(1)}]_{R,\, k}(t)
\\ \nonumber
&&\qquad+ \frac{1}{2}\left[c_{\delta,1}(t)-c_{\delta,3}(t)\right]\; [\d_i \delta^{(1)}\;\frac{\d^i}{\d^2}\theta^{(2)}]_{R,\, k}(t)
\eea
\bea \nonumber
&&\quad+\left[ \frac{1}{2} c_{\delta,1}(t)- c_{\delta,2}(t)+\frac{1}{2} c_{\delta,3}(t)\right]\;\\ \nonumber
&&\qquad \times\left[ [\d_i \delta^{(1)}\; \frac{\d_j\d^i}{\d^2}\theta^{(1)}\; \frac{\d^j}{\d^2}\theta^{(1)}]_{R,\, k}(t)+[\d_i\d_j \delta^{(1)}\;\frac{\d^i}{\d^2}\theta^{(1)}\frac{\d^j}{\d^2}\theta^{(1)}]_{R,\, k}(t)\right]+ 
\\ \nonumber
&&\qquad+c_{\delta^2,1}(t) \; [\delta^2]^{(2)}_{R,\, k}(t)+c_{\delta^2,2}(t) \; [\delta^2]^{(3)}_{R,\, k}(t)-2 \left[c_{\delta^2,1}(t)-c_{\delta^2,2}(t)\right][\delta^{(1)}\d_i \delta^{(1)}\frac{\d^i}{\d^2}\theta^{(1)}]_{R,\, k} 
\\ \nonumber
&& \qquad
+c_{s^2,1}(t) \; [s^2]_{R,\, k}^{(2)}(t)+c_{s^2,2}(t) \; [s^2]_{R,\, k}^{(3)}(t)-2 \left[c_{s^2,1}(t)-c_{s^2,2}(t)\right][s_{lm}^{(1)}\d_i (s^{lm})^{(1)}\frac{\d^i}{\d^2}\theta^{(1)}]_{R,\, k}
\\ \nonumber
&&\qquad +c_{st,1}(t) \; [st]^{(3)}_{R,\, k}(t)+ c_{\psi,1}(t)\;  \psi^{(3)}(k,t) + c_{\delta^3}[\delta^3]^{(3)}_{R,\, k}(t)+c_{\delta\,s^2}[\delta s^2]^{(3)}_{R,\, k}(t)\\ \nonumber
&&\qquad\quad+c_{s^3}[s^3]^{(3)}_{R,\, k}(t)+c_{\delta\,\epsilon^2}[\delta \epsilon^2]^{(3)}_{R,\, k}(t)\\ \nonumber
&&\qquad+ c_{\epsilon,1}(t) \;[\epsilon^{(1)}]_{R,\, k}+ c_{\epsilon,2}(t) \;[\epsilon^{(2)}]_{R,\, k}+\ldots\ ,
\eea
where for each operator ${\cal O}$ we have that
\be
[{\cal O}]_R(\vec x,t)={\cal O}(\vec x,t)-\langle{\cal O}(\vec x,t)\rangle- \alpha_{{\cal O}\,\delta,\rm UV} \delta^{(1)}(\vec x,t)+ \ldots\ ,
\ee
where $\ldots$ represent higher order terms, such as $\delta^{(2)}$ or stochastic terms $\epsilon$, that are unnecessary at one loop level, and $\alpha_{{\cal O},\rm UV}$ represents the UV sensitive part of the correlation function $\langle[{\cal O}]_k(t)\delta (k,t)\rangle'$. This can be easily read off from the expressions in~(\ref{eq:UVa}),(\ref{eq:UVb}),(\ref{eq:UVc}), and~(\ref{eq:UVd}). For example, for the operator $[\delta^{(1)}\delta^{(2)}](\vec x,t)$, we have that  $\alpha_{[\delta^{(1)}\delta^{(2)}]\delta^{(1)},\rm UV}=\frac{68}{21}\sigma^2(t)$. Notice that for all the operators of the form $[{\cal O}^{(1)}{\cal O}'{}^{(1)}](\vec x,t)$, we have that $\alpha_{[{\cal O}^{(1)}{\cal O}'{}^{(1)}]\delta,\rm UV}=0$. In the local-in-time approximation, one can put together all the perturbative terms to which an operator ${\cal O}$ contributes, and  can easily reconstruct the overall $\alpha_{{\cal O}\,\delta,\rm UV}$ for a given operator ${\cal O}$ when contracted with an operator $\delta$. For example, in this case one would have  $\alpha_{[\delta \theta]\delta,\rm UV}=\alpha_{[\delta^{(1)}\theta^{(2)}]\delta,\rm UV}+\alpha_{[\delta^{(2)}\theta^{(1)}]\delta,\rm UV}$. Again, as in the former case, one has to add a finite part to the $\alpha_{{\cal O}\,\delta^{(1)},\rm UV}$, no matter how this has been identified, to make the predictions agree with the observations.

This illustrates the renormalization procedure for the galaxies dark matter cross correlation. In summary, after doing this, in the non-local in time case, at one-loop, we are left with seven independent bias coefficients for the non-local in time treatment.
When we consider the local in time approximation, which, as discussed, is approximate but at some level applicable, the terms $a,\,b,\,c$ simplify relevantly, and we are left with only four bias coefficients. 

A comment on the higher derivative term. At the order at which we are working, this term has exactly the same functional form as the fourth term in~(\ref{eq:pointa}). The decision if this term should be included or not depends on the kind of observations we wish to do. The $c_s$ term that appears in correlation functions of dark matter is a physical term, in the sense that is related to observable quantities. For example it can be measured in lensing surveys. If however one is interested only in surveys of collapsed objects, where dark matter is not directly observed, then we can in principle redefine the $c_s$ for dark matter to absorb the coefficient~$c_{\d^2\delta,1}$. 

Finally, a comment of the stochastic bias counterterm. The only UV sensitive terms that we have identified at one-loop order and for the slope of the power spectrum that happens to be in our universe are of the form of the linear bias. However, one can imagine what would happen in different universes, where for example the power spectrum could be steeper. This is equivalent to foresee what will happen at sufficiently high loop order in perturbation theory with the slope $n\simeq -2.1$. It would then turn out that additional UV-sensitive terms are generated, requiring to shift new bias coefficients. In particular, now a divergent term can be generated for the higher derivative bias. It is less straightforward what happens to the terms in $c_{\delta^2,1}$ and $c_{s^2,1}$, which in our universe are not UV sensitive. In the limit $q\gg k$, we have that
\bea 
&&c_{\delta^2,1}(t) \; \int \frac{d^3q}{(2\pi)^3}\;  4\, F_{2,S}(\vec k-\vec q,\vec q)\;P_{11}(q;t,t)\,P_{11}(|\vec k-\vec q|;t,t)\\ \nonumber
&&\qquad\qquad\qquad\qquad\quad  \to\quad   -c_{\delta^2,1}(t) \frac{2}{21} k^2 \int \frac{d^3q}{(2\pi)^3}\;  \frac{1}{q^2}\;[P_{11}(q;t,t)]^2=-c_{\delta^2,1}(t) \frac{2}{21} \frac{k^2}{\knl^2}\frac{1}{\knl^3} \tilde\sigma^2(t)\  , 
\eea
where we have defined
\be
\tilde\sigma^2(t)=\knl^3 \int \frac{d^3q}{(2\pi)^3}\;  \frac{\knl^2}{q^2}\;[P_{11}(q;t,t)]^2 \ .
\ee
A similar behavior is also present for $c_{s^2,1}$.
If the slope $n$ of the power spectrum were to be higher than $-1/2$, this term would be UV sensitive. In this case, we would need to cancel this contribution. Since this term is not proportional to $P_{11}(k;t,t)$, the only available candidate is the stochastic counterterm. Indeed,  the fact that this term starts in $k^2$ tells us that it has the right functional form to be reabsorbed by a shift in the stochastic bias: 
\be
c_{\epsilon,1}(t)=c_{\epsilon,1,\rm finite}(t)+c_{\epsilon,1,\rm counter}(t)\ ,
\ee
with 
\be
c_{\epsilon,1,\rm counter}(t)=c_{\delta^2,1}(t)\frac{2}{21} \frac{1}{\gamma^{1/2}} \frac{\km^{3/2}}{\knl^{3/2}} \tilde\sigma^2(t)\ .
\ee

Alternatively, one could have re-defined the correlation function between $\epsilon_k$ and $[\d^2\Delta\tau]_k$ in the following way
\be
\langle\epsilon_k[\d^2\Delta\tau]_k\rangle=\frac{\gamma^{1/2}}{(\km^3\knl^3)^{1/2}}\cdot\frac{k^2}{\knl^2}+\frac{c_{\delta^2,1}(t)}{c_{\epsilon,1}(t)} \frac{2}{21} \frac{k^2}{\knl^2}\frac{1}{\knl^3} \tilde\sigma^2(t)\  , 
\ee
so that the overall contribution to the cross correlation with $k$-dependence of the form $k^2$ would fit observations~\footnote{Notice that these terms that are analytic in $k^2$, once Fourier transformed to real space, lead to terms that have support only at coincidence, that is that are  proportional to a $\delta$-function and derivatives of it. Their size is therefore  extremely sensitive to the procedure one uses to define correlation functions at vanishing distances. This ambiguity is mirrored in the EFT in the fact that we can redefine the size of the correlation functions of the stochastic terms so that we can always reabsorb the divergencies analytic in $k^2$.}.

This renormalization is not necessary in our universe at one-loop order, as the slope of the power spectrum near the non-linear scale is much less steep. It however gives us another interesting example of how renormalization works, and more importantly shows that a non vanishing correlation between the stochastic bias and the stochastic dark matter stress tensor is allowed by the symmetries of the problem, and it would indeed be necessary in universes with a very steep power spectrum close to the non-linear scale.

Our expression differs from the former treatment of~\cite{McDonald:2009dh} for a few important points. First, this is only the Eulerian expression that, as we argued, must be included in a Lagrangian calculation as we described in Sec.~\ref{sec:lagrangian}. Second, the expression is non-local in time, and if approximated as local in time, contains higher time derivative terms.  In particular in general this leads to a break up of the various terms that appear at each perturbative order in the dark matter correlation functions. The time-derivative terms that appear in the local-in-time approximation, as well as in general the spatial derivative, act relatively to the flux, not along the comoving coordinates. Third, the stochastic terms are treated differently. As we emphasized, some of these differences are irrelevant at one-loop order, but begin to play a role at higher loop order.

In App.~\ref{sec:galaxy-power} we give the same explicit formulas for the Galaxies Galaxies power spectrum at one-loop.

\section{Conclusions}
 
 We have discussed how to described correlation functions of collapsed objects in the EFTofLSS. We have started by formulating an Eulerian description of bias, where the overdensity of galaxies at a given location and time depends on the underlying tidal tensor and dark matter velocity gradients at the same location. We have then discussed how this number density depends on nearby points, performing a derivative expansion whose suppression is related to the mass of the object itself. We have then discussed that since the formation time of galaxies takes an Hubble time, their overdensity at a given time will depend not just on the dark matter properties at the same time, but on all earlier times up to order one Hubble time. This means that the EFT is non-local in time, as for the case of dark matter. Invariance under diffeomorphisms requires that the overdensity of galaxies at a given location depends on the dark matter field evaluated at earlier times and at the displaced location $\xfl(t,t')$ where the underlying dark matter fluid element used to be at those earlier times. We have discussed how the time non-locality can be dealt with: in practice it corresponds to having a different local bias not only for every field of which the galaxies are a tracer, but also for every order in perturbation theory at which that field is evaluated. This also offers a way to measure in simulations or observations the time non-locality of the theory: there is no need to measure the time-dependent kernels, but simply the different local bias coefficients associated to the same field when evaluated at different orders. 
 
In calculations for dark matter correlation functions, it has appeared that the local-in-time treatment seems to be a good approximation to the physics in our actual universe. We have given an argument for why this is the case: most of the modes that we integrate out and that affect the long-distance physics are characterized by a time scale quite faster than Hubble, though not by more than an order of magnitude. This justifies and explains why a local in time treatment seems to be a good approximation. We have discussed how in this case one has to introduce a dependence for the galaxy overdensity on time derivatives of fields, and that these time derivatives act along the flow.

We have then put into evidence how the Eulerian treatment expands not only in powers of the dark matter overdensities and short scale displacements, but also in long wavelength displacements that in our universe are too large to be treated perturbatively. This makes the Eulerian treatment non-convergent to the correct answer. This has lead us to formulate the theory of bias in Lagrangian space, according to which the overdensity of galaxies at a given location is determined by the initial configurations of the dark matter field at the location where that element of dark matter that ended up at that given location was initially. We have then slightly generalized formulas that were developed for performing calculations for dark matter correlation functions in Lagrangian space~\cite{Senatore:2014via}, to adapt them to the case of collapsed objects. In this way, we are able to upgrade Eulerian calculations to Lagrangian calculations in the same way as we did for dark matter. After upgrading the Eulerian calculation to a Lagrangian one, the resulting perturbative series is a series in powers of $k/\knl$ and $k/\km$, where $\knl$ is the wavenumber associated to the non-linear scale, and $\km$ is the wavenumber associated to the length scale from which the dark matter particles collapsed into a galaxy. The series is manifestly convergent, up to the scale where non-perturbative  corrections become important.
   
Having made Eulerian calculations useful again, we have given in detail the formulas for the dark matter galaxies cross correlations, and explicitly explained how the bias coefficients need to be renormalized in order to correctly parametrize any dependence on short distance physics that is not under perturbative control. 
   
The formulas and techniques we have provided and explained can be simply extended to the galaxies power spectra  and higher order correlation functions, such as the bispectrum, at one loop and beyond. 

One partial reason of concern is the fact that of order of a few bias parameters are in practice needed to describe galaxies correlation functions at high wavenumbers, with possible consequences for the predictivity of the theory. However, there are several reasons of hope: the same bias parameters appear in the several correlation functions, as for example the quadratic bias that enters at one loop in the power spectra enters also at tree level in the bispectrum; some of these bias parameters contribute in degenerate ways, so that the effective number of parameters gets reduced; the gain in sensitivity is extremely large as we reach higher wavenumbers with the theory, so that several parameters might be accommodated without too much of a loss. How much is the UV reach of the current formulations? How many parameters are needed and what is the remaining predictive power? These are very interesting questions that we leave to future work. The purpose of the the present work was to provide the formulas that should allow us to correctly answer those questions.

In the case of dark matter, the application of EFT techniques has showed the potential to increase the predictive reach of analytical techniques by a factor of about six in $k$-space, from about $k\simeq 0.1\hinvMpc$, where SPT fails, to $k\simeq 0.6 \hinvMpc$, implying the possible availability of a factor of 200 more modes amenable to analytic techniques in LSS surveys than previously believed. This could have potentially revolutionary consequences on what we can learn in Cosmology in the next decade. This results shows that there could be a lot to gain from using the EFTofLSS to describe the clustering of dark matter. It will be very interesting to compare how the formalism that we have developed here perform against simulations or data. We leave this to upcoming future work.

\subsubsection*{Acknowledgments}

We thank Matias Zaldarriaga for discussions. L.S. is supported by DOE Early Career Award DE-FG02-12ER41854 and by NSF grant PHY-1068380.

   \begin{appendix}
   
   \section*{Appendix}
   
   \section{\label{app:fluid_bias} Deriving the $\xfl$ terms}
   
   Here we explicitly derive the terms in the galaxies overdensity in (\ref{eq:euler_bias_4}) that are due to the Taylor expansion of $\xfl$. From eq.~(\ref{eq:xfl_expansion}) we have several terms. The first comes from taking the second term in~(\ref{eq:xfl_expansion}) and evaluating both terms at linear level. Keeping only the time-variables explicit for simplicity, we have
   \bea
  && -\int^t dt'\; H(t')\; \bar c_{\delta}(t,t')\;  \d_i \delta^{(1)}(t') \int^\tau_{\tau'} \d\tau''\; v^{(1)}{}^i(\tau'')=\\ \nonumber
  &&\quad=\int^t dt'\; H(t')\; \bar c_{\delta}(t,t')\;  \frac{D(t')}{D(t)}\d_i \delta^{(1)}(t) \int^\tau_{\tau'} \d\tau''\;\frac{D'(\tau'')}{D(\tau)}\frac{\d^i}{\d^2}\theta^{(1)} (\tau)\\ \nonumber
    &&\quad=\int^t dt'\; H(t')\; \bar c_{\delta}(t,t')\;  \frac{D(t')}{D(t)}\left[1-\frac{D(t')}{D(t)}\right]\d_i \delta^{(1)}(t) \frac{\d^i}{\d^2}\theta^{(1)} (t)\\ \nonumber
   &&\quad=\left[c_{\delta,1}(t)-c_{\delta,2}(t)\right]\d_i \delta^{(1)}(t) \frac{\d^i}{\d^2}\theta^{(1)} (t)\ .
   \eea
  From the same term, we can now take the term in $\delta$ at second order and in $v$ at first: $\d_i\delta^{(2)} \int \d\tau''\, v^{(1)}{}^i(\tau'')$. Trivially, this gives
     \bea
  && -\int^t dt'\; H(t')\; \bar c_{\delta}(t,t')\;  \d_i \delta^{(2)}(t') \int^\tau_{\tau'} \d\tau''\; v^{(1)}{}^i(\tau'')\\ \nonumber
&&  \quad =\left[c_{\delta,2}(t)-c_{\delta,3}(t)\right]\d_i \delta^{(2)}(t)  \frac{\d^i}{\d^2}\theta^{(1)} (t)\ .
   \eea
   The next option is to take, $\delta$ linear and $v$ at second order:  $\d_i\delta^{(1)} \int \d\tau''\, v^{(2)}{}^i(\tau'')$. We have
    \bea
  &&- \int^t dt'\; H(t')\; \bar c_{\delta}(t,t')\;  \d_i \delta^{(1)}(t') \int^\tau_{\tau'} \d\tau''\; v^{(2)}{}^i(\tau'')=\\ \nonumber
  &&\quad=\int^t dt'\; H(t')\; \bar c_{\delta}(t,t')\;  \frac{D(t')}{D(t)}\d_i \delta^{(1)}(t) \int^\tau_{\tau'} \d\tau''\;\frac{D(\tau'')D'(\tau'')}{D(\tau)^2}\frac{\d^i}{\d^2}\theta^{(2)}(\tau)\\ \nonumber
    &&\quad=\int^t dt'\; H(t')\; \bar c_{\delta}(t,t')\;  \frac{D(t')}{D(t)}\frac{1}{2}\left[1-\frac{D(t')^2}{D(t)^2}\right]\d_i \delta^{(1)}(t)  \frac{\d^i}{\d^2}\theta^{(2)} (t)\\ \nonumber
   &&\quad=\frac{1}{2}\left[c_{\delta,1}(t)-c_{\delta,3}(t)\right]\d_i \delta^{(1)}(t)  \frac{\d^i}{\d^2}\theta^{(2)} (t)\ .
   \eea
   Finally, in the cubic terms, at one-loop order we can take all the terms as linear. From the third term in (\ref{eq:xfl_expansion}), we have
  \bea
  &&- \frac{1}{2}\int^t dt'\; H(t')\; \bar c_{\delta}(t,t')\;  \d_i \delta^{(1)}(t') \int^\tau_{\tau'} d\tau''\; \d_jv^{(1)}{}^i(\tau'') \int^\tau_{\tau''} d\tau'''\; v^{(1)}{}^j(\tau''')\\ \nonumber
  &&\quad=- \frac{1}{2}\int^t dt'\; H(t')\; \bar c_{\delta}(t,t')\;  \frac{D(t')}{D(t)}\d_i \delta^{(1)}(t) \int^\tau_{\tau'} d\tau''\;\frac{D'(\tau'')}{D(\tau)} \int^\tau_{\tau''} d\tau'''\;\frac{D'(\tau'')}{D(\tau)}\frac{\d_j\d^i}{\d^2}\theta^{(1)} (\tau)\frac{\d^j}{\d^2}\theta^{(1)} (\tau)\\ \nonumber
    &&\quad=- \frac{1}{2}\int^t dt'\; H(t')\; \bar c_{\delta}(t,t')\;  \frac{D(t')}{D(t)}\left[\frac{1}{2}-\frac{D(t')}{D(t)}+\frac{1}{2}\frac{D(t')^2}{D(t)^2}\right]\d_i \delta^{(1)}(t)  \frac{\d^j\d^i}{\d^2}\theta^{(1)} (t) \frac{\d^j}{\d^2}\theta^{(1)}(t) \\ \nonumber
   &&\quad=-\frac{1}{2}\left[\frac{1}{2}c_{\delta,1}(t)-c_{\delta,2}(t)+\frac{1}{2}c_{\delta,3}(t)\right]\d_i \delta^{(1)}(t)  \frac{\d^i\d^j}{\d^2}\theta^{(1)} (t)\frac{\d^j}{\d^2}\theta^{(1)} (t)\ ,
   \eea
   and very similarly for the remaining cubic term.
   
   \subsection{Quasi-local approximation and the flow time-derivative\label{eq:flow-time-derivative}}
   
   Let us see how the time-derivative expansion emerges when we approximate the non-local in time bias kernels as quasi local. We will show how the time derivatives rearrange themselves in flow time derivatives. It is enough to focus on the first term, the others following in a similar way. 
   
 Let us therefore imagine that the time dependent bias kernels are non-zero only on a region of order  $t-t'\lesssim 1/\oms\ll 1/H$. In this case, starting from~(\ref{eq:euler_bias_3}), we can Taylor expand the fields around $t'\simeq t$ in the following way:
 \bea\label{eq:euler_bias_6}
&&\delta_M(\vec x,t)\supset \int^t dt'\; H(t')\;   \bar c_{\delta}(t,t')\; \delta(\xfl(t,t'),t') \\  \nonumber
&&\qquad=  \int^t d\Delta t\; H(t-\Delta t)\;   \bar c_{\delta}(t,t-\Delta t)\;  \delta(\xfl(t,t-\Delta t),t-\Delta t)\\ \nonumber
&&\qquad\simeq \int^t d\Delta t\; H(t-\Delta t)\;   \bar c_{\delta}(t,t-\Delta t)\;\left[\delta(\vec x,t)-\Delta t\left[\d_t\delta(\vec x,t)+v^i(\vec x, t)\d_i \delta(\vec x,t)\right]+\ldots\right]\\ \nonumber
&&\qquad =\left[ \int^t d\Delta t\; H(t-\Delta t)\;   \bar c_{\delta}(t,t-\Delta t)\right]\delta(\vec x,t) \\ \nonumber
&&\qquad\qquad+\left[ \int^t d\Delta t\; H(t-\Delta t)\;   \bar c_{\delta}(t,t-\Delta t)\, \Delta t\right]\left[\d_t\delta(\vec x,t)+v^i(\vec x, t)\d_i \delta(\vec x,t)\right]+\ldots\\ \nonumber
&&\qquad\equiv   c_{\delta}(t) \; \delta(\vec x,t)+   c_{\d_t\delta}(t)\; \frac{1}{\oms}\; \left[\d_t\delta(\vec x,t)+v^i(\vec x, t)\d_i \delta(\vec x,t)\right] +\ldots\ ,
 \eea
 where we have inserted a factor of $1/\oms$ on dimensional analysis so that $ c_{\delta}(t)$ and $ c_{\d_t\delta}(t)$ have comparable dimensionless sizes. Notice that in the third line we have the last term coming from the derivative of $\xfl(t,t')$ with respect to $t'$ evaluated at $t'=t$. This expansion converges if the time scale of the short modes that is encapsulated in $\bar c_\delta(t,t')$ is faster than the one of the long modes. We therefore see that upon assuming the quasi local in time approximation, the local in time bias contains terms with time derivatives that organize themselves as time-derivatives along the flow, as indeed it was dictated by diffeomorphisms invariance. This is what we wanted to show.
   
   \subsection{\label{app:approx_time} Leading Effects of the Quasi-Local Approximation}
   
   In sec.~\ref{sec:quasi-local}, we explained that the difference between the time-local approximation and the non-local in time treatment starts, for the case of dark matter, at about three-loop order. We here give an explicit explanation of this fact. At tree level, we can redefine $c_s$ to make the non-local in time treatment  equivalent to the local in time one. In formulas we can indeed define $c_{s,\rm local}$ as:
   \be\label{eq:cslocaldef}
   c_{s,\rm local}^2\; \delta^{(1)}(a,\vec x)=\int \frac{d a'}{a' \H'} \,K_{c_s}(a,a')\, \delta^{(1)}(a', \vec x)=  \left[\int \frac{d a'}{a' \H'} \,K_{c_s}(a,a')\frac{D(a')}{D(a)}\right]\, \delta^{(1)}(a,\vec  x)\ .
   \ee
   When we go to higher orders, we have to perform integrals of the following form
   \be
   \int \frac{d a'}{a' \H'} \,K_{c_s}(a,a')\, \delta^{(n)}(a', \vec x)=  \left[\int \frac{d a'}{a' \H'} \,K_{c_s}(a,a')\frac{D(a')^n}{D(a)^n}\right]\, \delta^{(n)}(a,\vec  x)\ ,
   \ee
with $n>1$. In the approximate time-local regime, the integrals can be manipulated to the following form
     \bea\nonumber
&&  \int \frac{d a'}{a' \H'} \,K_{c_s}(a,a')\frac{D(a')^n}{D(a)^n}=  \int \frac{d \Delta a}{(a-\Delta a)\H(a-\Delta a)} \,K_{c_s}(a,a-\Delta a)\frac{D(a-\Delta a)^n}{D(a)^n}  \\ \nonumber
  && \simeq \int \frac{d \Delta a}{(a-\Delta a)\H(a-\Delta a)} \,K_{c_s}(a,a-\Delta a) \frac{D(a-\Delta a)}{D(a)}\left(1-\Delta a\, (n-1)\, \frac{D'(a)}{D(a)}\right)\\ 
  &&=c_{s,\rm local}^2+\Delta c_{s,\rm local}^2 \frac{H}{\oms}\ .
   \eea  
 The last passage follows from recognizing that the first term in the integral is nothing but the same integral that defines $c_{s,\rm local}$ in~(\ref{eq:cslocaldef}), while the second term is a new contribution that, because the kernel has support for  $\Delta a/a\sim \frac{H}{\oms}$, is much smaller than the first one. Here $c_{s,\rm local}^2$ and $\Delta c_{s,\rm local}^2$ are expected to be comparable numbers. The leading terms in which such a treatment is applicable are $\delta^{(2)}$ and $\delta^{(3)}$. They give rise to the $c_s^2 P_\text{1-loop}$ terms, which contribute as two-loop terms. We therefore conclude that, as we wanted to show, up to two-loops in the dark matter power spectrum, we can treat the theory as local in time, without the addition of time-derivative operators; the mistake is of order of a two-loop term times $ \frac{H}{\oms}$, which makes it comparable to a three-loop term for the $k$'s of interest.

  Exactly the same discussion can be applied to galaxies. What makes the theory of dark matter quasi local up to two loops is that the leading term where a kernel appears is with $c_s^2$ term, which start contributing at one-loop order. For collapsed objects the bias kernels enter at tree level, which ensures that the theory can be treated as local in time and without any time derivative operator only up to one-loop order. 
     
 For completeness, we notice that   an alternative argument for the fact that for dark matter the time-non-locality start mattering at about three-loop order can be given in a way similar to the one we gave around~(\ref{eq:time_derivative}) for galaxies. If we assume that the theory is 	quasi-local in time, we need to ask when the first time-derivative operator starts contributing. For dark matter, one can include a leading time-derivative operator of the form
  \be
  c_{s,\text{ local},t}^2 \frac{1}{\oms}\frac{\d^2}{\knl^2}[\d_t\delta+v^i\d_i\delta]\ .
  \ee
  We can add a subtract $\frac{D'(a)}{D(a)}\delta$ and redefine  $ c_{s,\text{local}}\to  c_{s,\text{local}}+ c_{s,\text{ local},t}^2 \frac{\H}{\oms} \frac{D'(a)}{D(a)}$, which is a small correction, so that the time derivative operator starts as 
  \be
  c_{s,\text{ local},t}^2 \frac{1}{\oms}\frac{\d^2}{\knl^2}[\d_t\delta-\H\frac{D'(a)}{D(a)}\delta+v^i\d_i\delta]\ .
  \ee 
The resulting operator starts at quadratic level in the fluctuations, and because of the $\d^2/\knl^2$, it contributes as about a two-loop term times $H/\oms$, as we wanted to show.
     
  \section{Galaxies Galaxies Power Spectrum\label{sec:galaxy-power}}

Let us give explicit expressions for the Galaxies Galaxies power spectrum at one loop. We will see that several of the same bias coefficients that appear in the galaxies matter cross correlation  appear also here. Let us label the two populations of galaxies as $a$ and $b$. Similarly to the case of the cross power, we have several terms.

\begin{enumerate}

\item\label{point1p} The first term gives  
\bea\label{eq:point1p}
&&\langle\delta_{M_a}(k,t)\delta_{M_b}(k,t)\rangle'_a=\\ \nonumber
&&\qquad  \int^t dt'\; H(t')\int^t dt''\; H(t'')\; \left[   \bar c_{\delta,a}(t,t')\,\bar c_{\delta,b}(t,t'')\; \langle[\delta(\xfl,t')]_k[\delta(\xfl,t'')]_k\rangle'_\text{EFT-one-loop} \right]\\ \nonumber
&&\quad=c_{\delta,1,a}(t)\,c_{\delta,1,b}(t) \; \langle\delta^{(1)}(k,t)\delta^{(1)}(k,t)\rangle'+ c_{\delta,2,a}(t)\,c_{\delta,2,b}(t)\; \langle\delta^{(2)}(k,t)\delta^{(2)}(k,t)\rangle'\\ \nonumber
&&\quad+\left[c_{\delta,1,a}(t)-c_{\delta,2,a}(t)\right]\left[c_{\delta,1,b}(t)-c_{\delta,2,b}(t)\right]\langle[\d_i \delta^{(1)}\;\frac{\d^i}{\d^2}\theta^{(1)}]_k[\d_i \delta^{(1)}\;\frac{\d^i}{\d^2}\theta^{(1)}]_k\rangle'
\\
&&\quad+\left\{ c_{\delta,3,a}(t)\,c_{\delta,1,b}(t)\; \langle\delta^{(3)}(k,t)\delta^{(1)}(k,t)\rangle'\right.\\ \nonumber
&&\qquad+c_{\delta,3_{c_s},a}(t)\,c_{\delta,1,b}(t)\; \langle\delta^{(3)}_{c_s}(k,t)\delta^{(1)}(k,t)\rangle'\ \\ \nonumber
&&\qquad+ \left[c_{\delta,1,a}(t)-c_{\delta,2,a}(t)\right]\,c_{\delta,2,b}(t)\; \langle[\d_i \delta^{(1)}\; \frac{\d^i}{\d^2}\theta^{(1)}]_k(t)\delta^{(2)}(k,t)\rangle' \\ \nonumber
&&\quad\qquad+ \left[c_{\delta,2,a}(t)-c_{\delta,3,a}(t)\right]\,c_{\delta,1,b}(t)\; \langle[\d_i \delta^{(2)}\;\frac{\d^i}{\d^2}\theta^{(1)}]_k(t)\delta^{(1)}(k,t)\rangle'\\ \nonumber
&&\quad\qquad+ \frac{1}{2}\left[c_{\delta,1,a}(t)-c_{\delta,3,a}(t)\right]\,c_{\delta,1,b}(t)\; \langle[\d_i \delta^{(1)}\;\frac{\d^i}{\d^2}\theta^{(2)}]_k(t)\delta^{(1)}(k,t)\rangle'  \\ \nonumber
&&\qquad+\left[ \frac{1}{2} c_{\delta,1,a}(t)- c_{\delta,2,a}(t)+\frac{1}{2} c_{\delta,3,a}(t)\right]\,c_{\delta,1,b}(t)\;\\ \nonumber
&&\quad\qquad \times\left[ \langle[\d_i \delta^{(1)}\; \frac{\d_j\d^i}{\d^2}\theta^{(1)}\; \frac{\d^j}{\d^2}\theta^{(1)}]_k(t)\delta^{(1)}(k,t)\rangle'+\langle[\d_i\d_j \delta^{(1)}\;\frac{\d^i}{\d^2}\theta^{(1)}\frac{\d^j}{\d^2}\theta^{(1)}]_k(t)\delta^{(1)}(k,t)\rangle'\right] \\ \nonumber
&&\quad\left.+ a\leftrightarrow b\right\} \ .
\eea
In this term, no new bias coefficients appear than had not already appeared in the cross correlation. 

\item We then have 
\bea
&&\langle\delta_{M_a}(k,t)\delta_{M_b}(k,t)\rangle'_b=\quad  \int^t dt'\; H(t')\;  \int^t dt''\; H(t'')\\\ \nonumber
&&\times \left[   \bar c_{\delta^2,a}(t,t') \, \bar c_{\delta,b}(t,t'')\; \langle[\delta^2]_{k}(t')[\delta(\xfl,t'')]_k\rangle'+\bar c_{\delta,a}(t,t') \, \bar c_{\delta^2,b}(t,t'')\; \langle[\delta(\xfl,t')]_k[\delta^2]_k(t'')\rangle' \right]_\text{EFT-one-loop}\\ \nonumber
&&\quad  =\left\{c_{\delta^2,1,a}(t)\,c_{\delta,2,b}(t) \; \langle[\delta^2]^{(2)}_k(t)\delta^{(2)}(k,t)\rangle'+c_{\delta^2,2,a}(t)\,c_{\delta,1,b}(t)\; \langle[\delta^2]^{(3)}_k(t)\delta^{(1)}(k,t)\rangle'\right.\\ \nonumber
 &&\qquad\left.-2 \left[c_{\delta^2,1,a}(t)-c_{\delta^2,2,a}(t)\right]\,c_{\delta,1,b}(t)\langle[\delta^{(1)}\d_i \delta^{(1)}\frac{\d^i}{\d^2}\theta^{(1)}]_k(t)\delta^{(1)}(k,t)\rangle'\right.\\ \nonumber
 && \qquad\left.+\left[ c_{\delta,1,a}(t)-c_{\delta,2,a}(t)\right]c_{\delta^2,1,b}(t)\langle[\d_i\delta^{(1)}\frac{\d^j}{\d^2}\theta^{(1)}]_k[\delta^2]_k^{(2)}\rangle'+ a\leftrightarrow b\right\}\ ,
\eea
Again, we have no new bias coefficient.

\item The third term is very similar to the second one. It gives
\bea
&&\langle\delta_{M_a}(k,t)\delta_{M_b}(k,t)\rangle'_c=\quad  \int^t dt'\; H(t')\;  \int^t dt''\; H(t'')\\\ \nonumber
&&\times \left[   \bar c_{s^2,a}(t,t') \, \bar c_{\delta,b}(t,t'')\; \langle[s^2]_{k}(t')[\delta(\xfl,t'')]_k\rangle'+\bar c_{\delta,a}(t,t') \, \bar c_{s^2,b}(t,t'')\; \langle[\delta(\xfl,t')]+k[s^2]_k(t'')\rangle' \right]_\text{EFT-one-loop}\\ \nonumber
&&\quad  =\left\{c_{s^2,1,a}(t)\,c_{\delta,2,b}(t)\; \langle[s^2]^{(2)}_k(t)\delta^{(2)}(k,t)\rangle'+c_{s^2,2,a}(t)\,c_{\delta,1,b}(t) \; \langle[s^2]^{(3)}_k(t)\delta^{(1)}(k,t)\rangle'\right.\\ \nonumber
&&\qquad\left.-2 \left[c_{s^2,1}(t)-c_{s^2,2}(t)\right]\,c_{\delta,1,b}(t)\langle[s_{lm}^{(1)}\d_i (s^{lm})^{(1)}\frac{\d^i}{\d^2}\theta^{(1)}]_k(t)\delta^{(1)}(k,t)\rangle'\right.\\ \nonumber
 && \qquad\left.+\left[ c_{\delta,1,a}(t)-c_{\delta,2,a}(t)\right]c_{s^2,1,b}(t)\langle[\d_i\delta^{(1)}\frac{\d^j}{\d^2}\theta^{(1)}]_k[s^2]_k^{(2)}\rangle'+ a\leftrightarrow b\right\}\ ,
\eea
which gives us no new bias coefficient.

\item We can also contract the quadratic biases among themselves. We have
\bea\nonumber
&&\langle\delta_{M_a}(k,t)\delta_{M_b}(k,t)\rangle'_d= \ \   \int^t dt'\; H(t')\;  \int^t dt''\; H(t'')\; \left[   \bar c_{\delta^2,a}(t,t') \, \bar c_{\delta^2,b}(t,t'')\; \langle[\delta^2]_{k}(t')[\delta^2]_k(t'')\rangle'\right.\\  \nonumber
&&\qquad \left.+\bar c_{\delta^2,a}(t,t') \, \bar c_{s^2,b}(t,t'')\; \langle[\delta^2]_k(t')[s^2]_k(t'')\rangle'+\bar c_{s^2,a}(t,t') \, \bar c_{\delta^2,b}(t,t'')\; \langle[s^2]_k(t')[\delta^2]_k(t'')\rangle' \right.\\ \nonumber
&& \qquad \left. +\bar c_{s^2,a}(t,t') \, \bar c_{s^2,b}(t,t'')\; \langle[s^2]_k(t')[s^2]_k(t'')\rangle'\right]_\text{EFT-one-loop}\\ \nonumber
&&\qquad  =\left.c_{\delta^2,1,a}(t)\,c_{\delta^2,1,b}(t) \; \langle[\delta^2]^{(2)}_k(t)[\delta^2]_k^{(2)}(t)\rangle'+c_{s^2,1,a}(t)\,c_{s^2,1,b}(t) \; \langle[s^2]^{(2)}_k(t)[s^2]_k^{(2)}(t)\rangle'\right.\\ \nonumber
&&\qquad\quad+\left\{c_{\delta^2,1,a}(t)\,c_{s^2,1,b}(t) \; \langle[\delta^2]^{(2)}_k(t)[s^2]_k^{(2)}(t)\rangle'+ a\leftrightarrow b\right\}\ ,
\eea
which, again, gives us no new bias coefficient.

\item Let us now pass to the cubic biases. Similarly to the case of cross correlation, we have
\bea\nonumber
&&\langle\delta_{M_a}(k,t)\delta_{M_b}(k,t)\rangle'_e= \quad \int^t dt'\; H(t')\;\int^t dt''\; H(t'')\;\\ \nonumber
&&\qquad \times\, \left[   \bar c_{st,a}(t,t')\,\bar c_{\delta,b}(t,t'')\; \langle[st]_{k}(t')\delta(k,t'')\rangle'+ \bar c_{\delta,a}(t,t')\,\bar c_{st,b}(t,t'')\; \langle\delta(k,t')[st]_k(t'')\rangle'+\right.\\ \nonumber
&&\qquad\quad+   \bar c_{\psi,a}(t,t')\,\bar c_{\delta,b}(t,t'')\; \langle\psi(k,t')\delta(k,t'')\rangle'+ \bar c_{\delta,a}(t,t')\,\bar c_{\psi,b}(t,t'')\; \langle\delta(k,t')\psi(k,t'')\rangle'\\ \nonumber
&&\qquad  \quad+\bar c_{\delta^3,a}(t,t')\,\bar c_{\delta,b}(t,t'')\; \langle[\delta^3]_{k}(t')\delta(k,t)\rangle'+\bar c_{\delta,a}(t,t')\bar c_{\delta^3,b}(t,t'')\; \langle\delta(k,t')[\delta^3]_k(t'')\rangle'\\ \nonumber
&&\qquad\quad+\bar c_{\delta s^2,a}(t,t')\,\bar c_{\delta,b}(t,t'')\; \langle[\delta\,s^2]_{k}(t')\delta(k,t'')\rangle'+\bar c_{\delta,a}(t,t')\,\bar c_{\delta s^2,b}(t,t'')\; \langle\delta(k,t')[\delta\,s^2]_{k}(t'')\rangle'\\ \nonumber
&&\qquad\quad+\bar c_{\delta \epsilon^2,a}(t,t')\,\bar c_{\delta,b}(t,t'')\; \langle[\delta\, \epsilon^2]_{k}(t')\delta(k,t'')\rangle'+\bar c_{\delta,a}(t,t')\,\bar c_{\delta\epsilon^2,b}(t,t'')\; \langle\delta(k,t')[\delta\, \epsilon^2]_{k}(t'')\rangle'\\ \nonumber
&&\qquad\quad\left.+\bar c_{s^3,a}(t,t')\,\bar c_{\delta,b}(t,t'')\; \langle[s^3]_{k}(t')\delta(k,t'')\rangle'+\bar c_{\delta,b}(t,t')\,\bar c_{s^3,b}(t,t'')\; \langle\delta(k,t')[s^3]_{k}(t'')\rangle' \right]_\text{EFT-one-loop} \\ \nonumber
&& \qquad =\left\{c_{st,1,a}(t)\,c_{\delta,1,b}(t) \; \langle[st]_k^{(3)}(t)\delta^{(1)}(k,t)\rangle'+ c_{\psi,1,a}(t)c_{\delta,1,b}(t)\; \langle\psi^{(3)}(k,t)\delta^{(1)}(k,t)\rangle'\right.\\ \nonumber
&&\qquad\qquad+c_{\delta^3,1,a}(t) c_{\delta,1,b}(t) \; \langle[\delta^3]_k^{(3)}(t)\delta^{(1)}(k,t)\rangle'+c_{\delta s^2,1,a}(t)\, c_{\delta,1,b}(t)\; \langle[\delta\, s^2]_k^{(3)}(t)\delta^{(1)}(k,t)\rangle'\\ \nonumber 
&& \qquad\qquad+c_{\delta \epsilon^2,1,a}(t)\, c_{\delta,1,b}(t)\; \langle[\delta\, \epsilon^2]_k^{(3)}(t)\delta^{(1)}(k,t)\rangle'+c_{s^3,1,a}(t)\, c_{\delta,1,b}(t) \; \langle[s^3]_k^{(3)}(t)\delta^{(1)}(k,t)\rangle'\\ 
&&\qquad\qquad\left.+ a\leftrightarrow b\right\}\ ,
\eea
which gives us no new coefficients.

\item The leading spatial derivative term is also important at one-loop level.
\bea\nonumber
&&\langle\delta_{M_a}(k,t)\delta_{M_b}(k,t)\rangle'_f=  \int^t dt'\; H(t')\;\int^t dt''\; H(t'')\; \left[   \bar c_{\d^2\delta,a}(t,t') c_{\delta,b}(t,t'')    \;\frac{\d^2_{x_{\rm fl}}}{k_{M_a}^2}\langle\delta(\xfl,t')\delta(\vec x,t)\rangle'\right.\\ \nonumber
&&\qquad\qquad\qquad\qquad\qquad\qquad\left.+ \bar c_{\delta,a}(t,t') c_{\d^2 \delta,b}(t,t'')    \;\frac{\d^2_{x_{\rm fl}}}{k_{M_b}^2}\langle\delta(\xfl,t')\delta(\vec x,t)\rangle'\right]_\text{EFT-one-loop,\, k}  \\ 
&&\qquad \qquad =\left[\frac{k^2}{k_{M_a}^2}c_{\d^2\delta,1,a}(t) c_{\delta,1,b}(t)+\frac{k^2}{k_{M_b}^2}c_{\delta,1,a}(t) c_{\d^2\delta,1,b}(t) \right]\; \langle\delta^{(1)}(k,t)\delta^{(1)}(k,t)\rangle'\ , 
\eea
At the order at which we are working, this term has exactly the same functional form as the fourth term in~(\ref{eq:pointa}). We have already discussed the same situation in the dark matter galaxies cross correlation, and how this can lead to effectively one less bias parameter.

\item As for the cross correlation, we have the correlation of the stochastic bias with the stochastic dark matter stress tensor:
\bea
&&\langle\delta_{M_a}(k,t)\delta_{M_b}(k,t)\rangle'_g=c_{\epsilon,1,a}(t) \frac{k^2}{\knl^2} \langle[\epsilon_a]_k [\Delta\tau]_k\rangle'+c_{\epsilon,1,b}(t) \frac{k^2}{\knl^2} \langle [\Delta\tau]_k[\epsilon_b]_k\rangle'\\ \nonumber
&&\quad\qquad\qquad\qquad\qquad= c_{\epsilon,1,a}(t) \frac{k^2}{\knl^2} \frac{1}{\knl^{3/2}}\left[\frac{\gamma_a^{1/2}}{k_{M_a}^{3/2}}+\frac{\gamma_b^{1/2}}{k_{M_b}^{3/2}}\right]\ ,
\eea
whose coefficients have already appeared in the cross correlation.

\item In the galaxy power spectrum, we can also contract among themselves the stochastic biases. We have
\bea
&&\langle\delta_{M_a}(k,t)\delta_{M_b}(k,t)\rangle'_h= \quad \int^t dt'\; H(t')\;\int^t dt''\; H(t'')\;\\ \nonumber
&&\quad  \times\, \left[   \bar c_{\epsilon,a}(t,t')\,\bar c_{\epsilon,b}(t,t'')\; \langle\epsilon(k,t')\epsilon(k,t'')\rangle'+\right.\\ \nonumber
&&\qquad \left.+\bar c_{\epsilon \delta,a}(t,t')\,\bar c_{\epsilon \delta,b}(t,t'')\; \langle[\epsilon\delta]_{k}(t')[\epsilon\delta]_{k}(t'')\rangle'+\bar c_{\epsilon s,a}(t,t')\,\bar c_{\epsilon s,b}(t,t'')\; \langle[\epsilon s]_{k}(t')[\epsilon s]_{k}(t'')\rangle' \right.\\ \nonumber
&&\qquad \left.+\bar c_{\epsilon \delta,a}(t,t')\,\bar c_{\epsilon s,b}(t,t'')\; \langle[\epsilon\delta]_{k}(t')[\epsilon s]_{k}(t'')\rangle'+\bar c_{\epsilon s,a}(t,t')\,\bar c_{\epsilon \delta,b}(t,t'')\; \langle[\epsilon s ]_{k}(t')[\epsilon\delta]_{k}(t'')\rangle'\right. \\  \nonumber
&&\qquad \left.+\bar c_{\epsilon s,a}(t,t')\,\bar c_{\epsilon t,b}(t,t'')\; \langle[\epsilon s]_{k}(t')[\epsilon t]_{k}(t'')\rangle'+\bar c_{\epsilon t,a}(t,t')\,\bar c_{\epsilon s,b}(t,t'')\; \langle[\epsilon t ]_{k}(t')[\epsilon s]_{k}(t'')\rangle'\right. \\  \nonumber
&&\qquad \left.+\bar c_{\epsilon t,a}(t,t')\,\bar c_{\epsilon t,b}(t,t'')\; \langle[\epsilon t]_{k}(t')[\epsilon t]_{k}(t'')\rangle'\right]_\text{EFT-one-loop} \\  \nonumber
&& \quad  =c_{\epsilon,1,a}(t)\,c_{\epsilon,1,b}(t) \; \langle\epsilon^{(1)}(k,t)\epsilon^{(1)}(k,t)\rangle'+c_{\epsilon,2,a}(t)\,c_{\epsilon,2,b}(t) \; \langle\epsilon^{(2)}(k,t)\epsilon^{(2)}(k,t)\rangle'\\ \nonumber
&&\quad\qquad+ \left[c_{\epsilon,3,a}(t)\,c_{\epsilon,1,b}(t)+c_{\epsilon,1,a}(t)\,c_{\epsilon,3,b}(t)\right] \; \langle\epsilon^{(3)}(k,t)\epsilon^{(1)}(k,t)\rangle'\\ \nonumber
&&\quad \quad+ c_{\epsilon \delta,1,a}(t)\,c_{\epsilon \delta,1,b}(t) \; \langle[\epsilon \delta]_k^{(2)}(t)[\epsilon \delta]_k^{(2)}(t)\rangle'\\ \nonumber
&&\quad \quad+ c_{\epsilon s,1,a}(t)\,c_{\epsilon s,1,b}(t) \; \langle[\epsilon s]_k^{(2)}(t)[\epsilon s]_k^{(2)}(t)\rangle'\\ \nonumber
&&\quad \quad+\left[ c_{\epsilon s,1,a}(t)\,c_{\epsilon \delta,1,b}(t)+c_{\epsilon \delta,1,a}(t)\,c_{\epsilon s,1,b}(t)\right] \; \langle[\epsilon \delta]_k^{(2)}(t)[\epsilon s]_k^{(2)}(t)\rangle'\\ \nonumber
&&\quad \quad+\left[ c_{\epsilon s,1,a}(t)\,c_{\epsilon t,1,b}(t)+c_{\epsilon t,1,a}(t)\,c_{\epsilon s,1,b}(t)\right] \; \langle[\epsilon s]_k^{(2)}(t)[\epsilon t]_k^{(2)}(t)\rangle'\\ \nonumber
&&\quad \quad+c_{\epsilon t,1,a}(t)\,c_{\epsilon t,1,b}(t)\; \langle[\epsilon t]_k^{(2)}(t)[\epsilon t]_k^{(2)}(t)\rangle'\ ,
\eea
where we have used that contractions such as
\be
\langle[\epsilon t]_k^{(2)}(t)[\epsilon \delta]_k^{(2)}(t)\rangle'\sim\langle \epsilon_{ij} \epsilon\rangle'\langle t^{ij}\delta\rangle'\propto\delta_{ij} \langle t^{ij} \delta\rangle'=0\ .
\ee

Here we see that new bias coefficients that had not appeared at the same order in the cross correlation find a role.

As we discussed in the main text, in order to estimate the contribution of the stochastic terms, one needs to be careful. In fact, since $\langle[\epsilon_{a}]_k[\epsilon_{b}]_k\rangle'\sim \gamma_{ab} (\km_a^3\km_b^3)^{-1/2}$, the perturbative series in the stochastic terms starts at very low order in the derivative expansion. However, some of the induced terms are analytic in $k^2$, which means that, when comparing with observations or simulations, they are very sensitive to the definition of the correlation functions at coincidence.

Following the same steps as in the case of the galaxies matter cross power, one can derive the exact expressions, work out the  renormalization, and the simplified expressions for the local in time approximation. This follows the same easy steps as we described earlier. We leave this for future work when we plan to compare the results of the EFT with numerical simulations.  We simply mention the most novel fact with respect the dark matter galaxies cross correlation. Here the terms such as the ones in $d$ as well as some others are UV sensitive and lead to contributions that are $k$ independent, scaling as $k^0$ and with no power of $P_{11}(k;t,t)$. They can be reabsorbed by a redefinition of the correlation function of the stochastic bias. Notice that one of the two ways to renormalize the cross correlation that we explained was to shift the coefficient of the stochastic bias $c_\epsilon$.  So, if we had chosen this procedure, there is no more freedom in fixing that coefficient. But there was still an ambiguity in what we meant by the correlation of the stochastic terms. We have:
\be
\langle\delta_{M_a}(k,t)\delta_{M_b}(k,t)\rangle'_h= c_{\epsilon,1,a}c_{\epsilon,1,b}\langle[\epsilon_{a}]_k[\epsilon_{b}]_k\rangle'=c_{\epsilon,1,a}c_{\epsilon,1,b} \frac{\gamma_{ab}}{(\km_a^3\km_b^3)^{1/2}}  \ ,
\ee
and one has the freedom to shift the parameter $\gamma_{ab}$, allowing to absorb the UV dependence. Equivalently, one could have kept fixed $c_{\epsilon,1,a}$ and decided to renormalize as independent the power spectra of the stochastic bias terms and their cross correlation with the stochastic stress tensor of dark matter. This is the second method that we described to renormalize this kind of analytic terms in the cross correlation. In a sense, this is like saying that we need to renormalize the stochastic bias operator.

\end{enumerate}

   \end{appendix}
   
 \begingroup\raggedright\endgroup


\begin{thebibliography}{10}

\bibitem{Baumann:2010tm} 
  D.~Baumann, A.~Nicolis, L.~Senatore and M.~Zaldarriaga,
  ``Cosmological Non-Linearities as an Effective Fluid,''
  JCAP {\bf 1207}, 051 (2012)
  [arXiv:1004.2488 [astro-ph.CO]].




\bibitem{Carrasco:2012cv} 
  J.~J.~M.~Carrasco, M.~P.~Hertzberg and L.~Senatore,
  ``The Effective Field Theory of Cosmological Large Scale Structures,''
  JHEP {\bf 1209}, 082 (2012)
  [arXiv:1206.2926 [astro-ph.CO]].


\bibitem{Porto:2013qua}
  R.~A.~Porto, L.~Senatore and M.~Zaldarriaga,
  ``The Lagrangian-space Effective Field Theory of Large Scale Structures,''
  arXiv:1311.2168 [astro-ph.CO].


\bibitem{Senatore:2014via}
  L.~Senatore and M.~Zaldarriaga,
  ``The IR-resummed Effective Field Theory of Large Scale Structures,''
  arXiv:1404.5954 [astro-ph.CO].




\bibitem{Carrasco:2013mua}
  J.~J.~M.~Carrasco, S.~Foreman, D.~Green and L.~Senatore,
  ``The Effective Field Theory of Large Scale Structures at Two Loops,''
  arXiv:1310.0464 [astro-ph.CO].


\bibitem{Angulo:2014tfa}
  R.~E.~Angulo, S.~Foreman, M.~Schmittfull and L.~Senatore,
  ``The One-Loop Matter Bispectrum in the Effective Field Theory of Large Scale Structures,''
  arXiv:1406.4143 [astro-ph.CO].

\bibitem{Baldauf:2014qfa}
  T.~Baldauf, L.~Mercolli, M.~Mirbabayi and E.~Pajer,
  ``The Bispectrum in the Effective Field Theory of Large Scale Structure,''
  arXiv:1406.4135 [astro-ph.CO].

\bibitem{Crocce:2005xy}
  M.~Crocce and R.~Scoccimarro,
  ``Renormalized cosmological perturbation theory,''
  Phys.\ Rev.\ D {\bf 73} (2006) 063519
  [astro-ph/0509418].



\bibitem{McDonald:2009dh}
  P.~McDonald and A.~Roy,
  ``Clustering of dark matter tracers: generalizing bias for the coming era of precision LSS,''
  JCAP {\bf 0908} (2009) 020
  [arXiv:0902.0991 [astro-ph.CO]].


\bibitem{Smith:2006ne}
  R.~E.~Smith, R.~Scoccimarro and R.~K.~Sheth,
  ``The Scale Dependence of Halo and Galaxy Bias: Effects in Real Space,''
  Phys.\ Rev.\ D {\bf 75} (2007) 063512
  [astro-ph/0609547].


\bibitem{Matsubara:1999qq}
  T.~Matsubara,
  ``Stochasticity of bias and nonlocality of galaxy formation: Linear scales,''
  Astrophys.\ J.\  {\bf 525} (1999) 543
  [astro-ph/9906029].

\bibitem{Matsubara:2011ck}
  T.~Matsubara,
  ``Nonlinear Perturbation Theory Integrated with Nonlocal Bias, Redshift-space Distortions, and Primordial Non-Gaussianity,''
  Phys.\ Rev.\ D {\bf 83} (2011) 083518
  [arXiv:1102.4619 [astro-ph.CO]].

\bibitem{Kehagias:2013rpa}
  A.~Kehagias, J.~Nore–a, H.~Perrier and A.~Riotto,
  ``Consequences of Symmetries and Consistency Relations in the Large-Scale Structure of the Universe for Non-local bias and Modified Gravity,''
  Nucl.\ Phys.\ B {\bf 883} (2014) 83
  [arXiv:1311.0786 [astro-ph.CO]].



\bibitem{Cheung:2007st}
  C.~Cheung, P.~Creminelli, A.~L.~Fitzpatrick, J.~Kaplan and L.~Senatore,
  ``The Effective Field Theory of Inflation,''
  JHEP {\bf 0803} (2008) 014
  [arXiv:0709.0293 [hep-th]].


\bibitem{Senatore:2010wk}
  L.~Senatore and M.~Zaldarriaga,
  ``The Effective Field Theory of Multifield Inflation,''
  JHEP {\bf 1204} (2012) 024
  [arXiv:1009.2093 [hep-th]].


\bibitem{Carroll:2013oxa}
  S.~M.~Carroll, S.~Leichenauer and J.~Pollack,
  ``A Consistent Effective Theory of Long-Wavelength Cosmological Perturbations,''
  arXiv:1310.2920 [hep-th].


\bibitem{Matsubara:2008wx}
  T.~Matsubara,
  ``Nonlinear perturbation theory with halo bias and redshift-space distortions via the Lagrangian picture,''
  Phys.\ Rev.\ D {\bf 78} (2008) 083519
   [Erratum-ibid.\ D {\bf 78} (2008) 109901]
  [arXiv:0807.1733 [astro-ph]].




\bibitem{Senatore:2012ya}
  L.~Senatore and M.~Zaldarriaga,
  ``The constancy of $\zeta$ in single-clock Inflation at all loops,''
  JHEP {\bf 1309} (2013) 148
  [arXiv:1210.6048 [hep-th]].


\bibitem{Carlson:2012bu} 
  J.~Carlson, B.~Reid and M.~White,
  ``Convolution Lagrangian perturbation theory for biased tracers,''
  arXiv:1209.0780 [astro-ph.CO].

\bibitem{Biagetti:2014pha}
  M.~Biagetti, V.~Desjacques, A.~Kehagias and A.~Riotto,
  ``Non-local halo bias with and without massive neutrinos,''
  Phys.\ Rev.\ D {\bf 90} (2014) 045022
  [arXiv:1405.1435 [astro-ph.CO]].



\bibitem{Pajer:2013jj} 
  E.~Pajer and M.~Zaldarriaga,
  ``On the Renormalization of the Effective Field Theory of Large Scale Structures,''
  arXiv:1301.7182 [astro-ph.CO].


\bibitem{Mercolli:2013bsa}
  L.~Mercolli and E.~Pajer,
  ``On the Velocity in the Effective Field Theory of Large Scale Structures,''
  arXiv:1307.3220v1  [astro-ph.CO].



\bibitem{Assassi:2014fva}
  V.~Assassi, D.~Baumann, D.~Green and M.~Zaldarriaga,
  ``Renormalized Halo Bias,''
  arXiv:1402.5916 [astro-ph.CO].




\end{thebibliography}
\end{document}